\documentclass[a4paper,11pt]{article}
\usepackage{a4wide,amsmath,amssymb}
\usepackage[english]{babel}

\usepackage{slashed}

\csname@addtoreset\endcsname{equation}{section}

\date{}
\begin{document}
\begin{titlepage}
\title{
\begin{flushright}
{\small MIFPA-11-50}
~\\
~\\
\end{flushright}
{\bf More on integrable structures of superstrings in $AdS_4\times CP^3$ and $AdS_2\times S^2\times T^6$ superbackgrounds}
~\\
\medskip
\medskip
\medskip
\author{Alessandra Cagnazzo$^{a,b}$, Dmitri Sorokin$^{b,c}$ and Linus Wulff$^d$
~\\
~\\
{\small $^a$\it Dipartimento di Fisica ``Galileo Galilei", Universit\'a degli Studi di Padova}
~\\
\small{\it and}
~\\
{\small $^b$ \it INFN, Sezione di Padova, via F. Marzolo 8, 35131 Padova, Italia}
~\\
~\\
{\small $^c$ \it Department of Theoretical Physics, the University of the Basque Country UPV/EHU,}\\
{\small \it P.O. Box 644, 48080 Bilbao, Spain}
~\\
\small{\it and}
~\\
{\small IKERBASQUE, \it Basque Foundation for Science, 48011, Bilbao, Spain}
~\\
~\\
\small{$^d$\it George P. \& Cynthia Woods Mitchell Institute}
{\small\it for Fundamental Physics and Astronomy,}
~\\
{\small\it Texas A\&M University, College Station, TX 77843, USA}
}
 }
\maketitle

\begin{abstract}
In this paper we continue the study, initiated in \cite{Sorokin:2010wn,Sorokin:2011rr}, of the classical integrability of Green--Schwarz superstrings in $AdS_4\times CP^3$ and $AdS_2\times S^2\times T^6$ superbackgrounds whose spectrum contains non--supercoset worldsheet degrees of freedom corresponding to broken supersymmetries in the bulk. We derive an explicit expression, to all orders in the coset fermions and to second order in the non--coset fermions, which extends the supercoset Lax connection in these backgrounds with terms depending on the non--coset fermions. An important property of the obtained form of the Lax connection is that it is invariant under $\mathbb Z_4$--transformations of the superisometry generators and the spectral parameter. This demonstrates that the contribution of the non--coset fermions does not spoil the $\mathbb Z_4$--symmetry of the super--coset Lax connection which is of crucial importance for the application of Bethe--ansatz techniques. The expressions describing the $AdS_4\times CP^3$ and $AdS_2\times S^2\times T^6$ superstring sigma--models and their Lax connections have a very similar form. This is because their amount of target--space supersymmetries complement each other to $32=24+8$, the maximal number of 10d type II supersymmetries. As a byproduct, this similarity has allowed us to obtain the form of the geometry of the complete type IIA $AdS_2\times S^2\times T^6$ superspace to all orders in the \emph{coset} fermions and to the second order in the \emph{non--coset} ones.

\end{abstract}

\thispagestyle{empty}
\end{titlepage}

\tableofcontents

\section{Introduction}
In this paper we continue the study of the classical integrability of Green--Schwarz superstrings in $AdS_4\times CP^3$ and $AdS_2\times S^2\times T^6$ superbackgrounds, whose spectrum contains non--supercoset worldsheet degrees of freedom, initiated in \cite{Sorokin:2010wn,Sorokin:2011rr}.

The integrability properties of superstrings on semi--symmetric coset superspaces $G/H$ with $\mathbb Z_4$--grading are, by now, very well understood. The prescription for constructing a Lax representation of the equations of motion of $2d$ sigma--models on the supercoset $G/H$ (that generates an infinite set of conserved charges) has been proposed in \cite{Bena:2003wd} and applied to various concrete examples \cite{Adam:2007ws} including the maximally supersymmetric type IIB $AdS_5\times S^5$ superstring whose target superspace is $\frac{PSU(2,2|4)}{SO(1,4)\times SO(5)}$ and an $\frac{OSp(6|4)}{SO(1,3)\times U(3)}$ sigma--model \cite{Arutyunov:2008if,Stefanski:2008ik} which is a kappa--symmetry gauge--fixed sub--sector of the Green--Schwarz superstring on a type IIA $AdS_4\times CP^3$ superspace \cite{Gomis:2008jt}. Other examples of interest, in particular in the AdS/CFT context, are string sigma--models on $\frac{PSU(1,1|2)\times PSU(1,1|2)}{SU(1,1)\times SU(2)}$ whose bosonic body is the $6d$ symmetric space $AdS_3\times S^3$ and on $\frac{D(2,1;\alpha)\times D(2,1;\alpha)}{SO(1,2)\times SO(3)\times SO(3)}$ having $AdS_3\times S^3\times S^3$ as its bosonic subspace \cite{Babichenko:2009dk}.
These cases are related to $10d$ superstrings compactified on $AdS_3\times S^3\times M_4$ (where $M_4$ is $T^4$ or $S^3\times S^1$) that preserve 16 target--space supersymmetries.
 Another example, which we will consider here, is a superstring on the coset superspace $\frac{PSU(1,1|2)}{SO(1,1)\times U(1)}$ with the $4d$ bosonic subspace $AdS_2\times S^2$ and eight supersymmetric Grassmann--odd directions. This model is a consistent truncation of a $10d$ Green--Schwarz superstring on $AdS_2\times S^2\times T^6$ or $AdS_2\times S^2\times CY^3$ (see \cite{Sorokin:2011rr} for more details and references). It is useful to have a supercoset description which captures the full 10d bosonic geometry of $AdS_2\times S^2\times T^6$ rather than a truncation to 4d. This can be achieved by noting that $AdS_2\times S^2\times\mathbb R^6$, with eight fermionic directions, is described by the supercoset $\frac{PSU(1,1|2)\rtimes E(6)}{SO(1,1)\times U(1)\times SO(6)}$, where the semi-direct product with $E(6)$, the Euclidean group in six dimensions, accounts for the $\mathbb R^6$ factor. Since $AdS_2\times S^2\times T^6$ is locally the same as $AdS_2\times S^2\times\mathbb R^6$, and we will only be interested in the local geometry, this gives us a (local) supercoset description of $AdS_2\times S^2\times T^6$.

Among the above examples only the $AdS_5\times S^5$ superstring is maximally supersymmetric in the $10d$ target space. Its number of supersymmetries and corresponding string fermionic modes is 32 coinciding with the number of  Grassmann--odd directions of $\frac{PSU(2,2|4)}{SO(1,4)\times SO(5)}$. In other words, all the worldsheet fermionic modes of the $AdS_5\times S^5$ string are in one to one correspondence with the Grassmann directions of the supercoset space which fully describes the supergeometry of the type IIB $AdS_5\times S^5$ supergravity solution. As a consequence, the prescription of \cite{Bena:2003wd} for the construction of a zero--curvature Lax connection from the $\mathbb Z_4$--graded components of the Cartan form on $\frac{PSU(2,2|4)}{SO(1,4)\times SO(5)}$ demonstrates the classical integrability of the full Green--Schwarz superstring in the  $AdS_5\times S^5$ superbackground which coincides with $\frac{PSU(2,2|4)}{SO(1,4)\times SO(5)}$.

Other, less supersymmetric, cases turn out to be more involved. For instance, the Green--Schwarz superstring on $AdS_4\times CP^3$ is invariant under 24 target--space supersymmetries that generate the superisometry group $OSp(6|4)$. The type IIA superspace, in which the string moves, has 32 fermionic directions while the supercoset $\frac{OSp(6|4)}{SO(1,3)\times U(3)}$ only has 24. This means that only 24 of the 32 fermionic modes on the string worldsheet can be associated with the supercoset Grassmann--odd directions, while the 8 remaining fermionic modes (corresponding to broken target--space supersymmetries) do not have this group--theoretical meaning. In fact, the complete type IIA $AdS_4\times CP^3$ superspace is not a supercoset, though it has the $OSp(6|4)$ isometries. Its geometry is much more complicated and reduces to that of $\frac{OSp(6|4)}{SO(1,3)\times U(3)}$ only in the sub--superspace in which the 8 non--supersymmetric fermionic coordinates are put to zero \cite{Gomis:2008jt}. In the Green--Schwarz superstring sigma--model on $AdS_4\times CP^3$ superspace these eight non--supercoset fermionic modes can be put to zero by partially gauge fixing the kappa--symmetry  for almost all classical configurations of the string. This however is not possible when the string motion is restricted to the $AdS_4$ subspace \cite{Arutyunov:2008if,Gomis:2008jt} or when the string forms a worldsheet instanton by wrapping a $CP^1$ cycle in $CP^3$ \cite{Cagnazzo:2009zh}. In these cases the supercoset kappa--symmetry gauge is inadmissible, and the non--coset fermions carry physical worldsheet degrees of freedom.\footnote{Subtleties of gauge fixing kappa-symmetry in a way consistent with the light--cone gauge in a near plane--wave limit of $AdS_4 \times CP^3$ has been discussed in \cite{Astolfi:2009qh}. } As a result, the construction of a Lax connection of the Green--Schwarz superstring in the full $AdS_4\times CP^3$ superspace, in general, should include the contribution of the non--coset fermions which will thus modify the form of the supercoset Lax connection of \cite{Arutyunov:2008if,Stefanski:2008ik} by terms whose structure is not captured by the prescription of \cite{Bena:2003wd}.

The situation becomes even more interesting and complicated in less supersymmetric cases such as strings on $AdS_3\times S^3\times M_4$ and $AdS_2\times S^2\times T^6$. For instance, as we have already mentioned, in $AdS_2\times S^2\times T^6$ only 8 target--space supersymmetries corresponding to the Grassmann--odd directions of  $\frac{PSU(1,1|2)\rtimes E(6)}{SO(1,1)\times U(1)\times SO(6)}$ are preserved and hence the other 24 fermionic modes of the Green--Schwarz superstring cannot be associated with the supercoset. Moreover, since there are only 16 kappa--symmetries, they can gauge away not more than 16 of these fermions, so that at least 8 of the non--coset worldsheet fermions carry physical degrees of freedom and will always contribute to the structure of the Lax connection of the complete $10d$ theory.

To deal with the non--coset fermions, an alternative prescription for constructing Lax connections has been proposed in \cite{Sorokin:2010wn}. It uses the Noether currents of the isometries of the (super)background as building blocks of the Lax connection and can thus be applied to more general cases than the $G/H$ sigma--models with $\mathbb Z_4$--grading. Using this procedure, zero--curvature Lax connections for superstrings on $AdS_4\times CP^3$ and $AdS_2\times S^2\times T^6$ have been constructed up to second order in the 32 fermionic modes, respectively, in \cite{Sorokin:2010wn} and \cite{Sorokin:2011rr}. In addition, in \cite{Sorokin:2010wn} a Lax connection to all orders in non--coset fermions has been constructed in a special kappa--symmetry gauge of \cite{Grassi:2009yj} in the sub--sector of the $AdS_4\times CP^3$ superstring which cannot be reduced to the $\frac{OSp(6|4)}{SO(1,3)\times U(3)}$ supercoset, thus providing evidence for the classical integrability of the complete theory.

When the non--coset fermions are put to zero, the Lax connections of \cite{Sorokin:2010wn,Sorokin:2011rr} are related to those of
 \cite{Bena:2003wd,Adam:2007ws,Arutyunov:2008if,Stefanski:2008ik} (truncated to the second order in the \emph{coset} fermions) by a superisometry gauge transformation that depends on the spectral parameter \cite{Sorokin:2010wn}.  To understand how the presence of the \emph{non--coset} fermions modifies e.g. the algebraic curve constructed with the use of the $\mathbb Z_4$--graded supercoset Lax connection and, hopefully, to reveal a role of the non--coset fermionic and bosonic modes in the corresponding Bethe--ansatz techniques, it seems useful to have at hand an explicit expression which demonstrates how the $\mathbb Z_4$--graded supercoset Lax connection gets generalized by terms depending on the non--coset fermions. In this paper we provide such an expression for the Lax connections of the superstring on $AdS_4\times CP^3$ and $AdS_2\times S^2\times T^6$ to all orders in the coset fermions and up to the second order in the non--coset fermions. Interestingly enough, the Lax connections of the $AdS_4\times CP^3$ and $AdS_2\times S^2\times T^6$ superstrings have formally a very similar form. This is because their numbers of target--space supersymmetries complement each other to $32=24+8$, the maximal number of \emph{10d} type II supersymmetries, and the projectors which split 32--component fermions into 24-- and 8--component ones are the same in both of the cases.

 This similarity actually has helped us to guess the form of the  $AdS_2\times S^2\times T^6$ Lax connection upon having constructed the  $AdS_4\times CP^3$ one using the knowledge of the complete $AdS_4\times CP^3$ supergeometry and the superstring equations of motion. As a byproduct, this has also allowed us to get corrections due to the non--coset fermions to the geometry of the $\frac{PSU(1,1|2)\rtimes E(6)}{SO(1,1)\times U(1)\times SO(6)}$ supercoset thus obtaining the form of the geometry of the complete type IIA $AdS_2\times S^2\times T^6$ superspace to all orders in the \emph{coset} fermions and to the second order in the \emph{non--coset} ones. These results make explicit the general discussion of \cite{Sorokin:2011rr} about the structure of the  $AdS_2\times S^2\times T^6$ supergeometry which ensures that the $\frac{PSU(1,1|2)}{SO(1,1)\times U(1)}$ sigma--model is a consistent truncation of the complete $10d$ superstring action on $AdS_2\times S^2\times T^6$.

 An important property of the obtained form of the Lax connection is that it is invariant under the $\mathbb Z_4$ transformations of the superisometry generators provided that the spectral parameter \textbf{x} gets replaced with its inverse $\frac{1}{\mathbf x}$. This demonstrates that the contribution of the non--coset fermions does not spoil the $\mathbb Z_4$--symmetry of the \emph{supercoset} Lax connection which is of crucial importance for the Bethe ansatz equations, both classical and quantum \cite{Beisert:2005bm,Gromov:2008bz,Babichenko:2009dk,SchaferNameki:2010jy}.

The paper is organized as follows.

In Section 2 we explain our conventions and notation and describe some general properties of the supercoset Lax connection, and the relation of its zero curvature condition to the equations of motion of the corresponding $2d$ dynamical system.

In Section 3 we extend the supercoset Lax connection with contributions coming from the string fermionic modes associated with broken targetspace supersymmetries. In Section \ref{AdS4} we sketch the construction of the Lax connection of the $AdS_4\times CP^3$ superstring starting from that of the $\frac{OSp(6|4)}{SO(1,3)\times U(3)}$ sigma--model and modifying it with terms containing the non--coset fermions (which we call $\upsilon$) in such a way that the zero--curvature condition is satisfied order by order in $\upsilon$ if the worldsheet fields obey the superstring equations of motion.

In Section \ref{AdS2} we pass to the consideration of the $AdS_2\times S^2\times T^6$ case and assume that its Lax connection has a similar form to that of the $AdS_4\times CP^3$ superstring but with the role of the coset and non--coset fermions interchanged and with an appropriate redefinition of the form of the gamma--matrices involved in the construction. We then require that this Lax connection has zero curvature, derive from this condition the equations of motion of the $AdS_2\times S^2\times T^6$ superstring  and reconstruct the geometry of its target superspace. Namely, we find the bosonic and fermionic vielbeins, the spin connection and the NS--NS three--form superfield strength of the $AdS_2\times S^2\times T^6$ superspace to all orders in the fermions parametrizing the supercoset $\frac{PSU(1,1|2)\rtimes E(6)}{SO(1,1)\times U(1)\times SO(6)}$ and to the second order in the non--coset fermions.

In Section 4 we demonstrate that the obtained Lax connections are invariant under the $\mathbb Z_4$--transformations and discuss their relation to conserved currents and the Lax connections constructed in \cite{Sorokin:2010wn,Sorokin:2011rr}.

In the Conclusions we discuss open problems, possible generalizations and applications of the results obtained.

\section{Setting the stage}
\subsection{Main notation and conventions}
\label{sec:notation}
We use the  metric with the `mostly plus' signature $(-,+,\cdots,+)$.
Generically, the tangent space vector indices are labelled by letters from the beginning of the
Latin alphabet,  while  letters from the middle of the Latin alphabet stand for curved (world)
indices. The spinor indices are labelled by Greek letters from the beginning of the alphabet, while their curved (world) counterparts are denoted by letters from the middle of the Greek alphabet.

The bosonic coordinates of the ten--dimensional type IIA target superspace in which the string moves are denoted by $X^M$ $(M=0,1,\cdots,9)$ and the Grassmann--odd coordinates are denoted by $\Theta^{\mu}$ ($\mu=1,\cdots, 32)$. Since we consider the string sigma--model we shall always assume that $X^M$ and $\Theta^{\mu}$ depend on the string worldsheet variables $\xi^i=(\tau,\sigma)$.

The geometry of the target superspace is encoded in the form  of the vector $\mathcal E^A(X,\Theta)$ and spinor $\mathcal E^\alpha(X,\Theta)$ supervielbeins, and spin connection $\Omega^{AB}(X,\Theta)$. In the string sigma--model these one--forms are pulled back on the string worldsheet, which will always be implicit in what follows, \emph{e.g.} $\mathcal E^A(X,\Theta)=d\xi^i(\partial_iX^M\mathcal E_M{}^A(X,\Theta)+\partial_i\Theta^\mu\mathcal E_\mu{}^A(X,\Theta))$. The $10d$ supergeometry is subject to the basic torsion constraint which we choose to be
\begin{eqnarray}\label{TA}
T^A\equiv d\mathcal E^A+\mathcal E^B\Omega_B{}^A=-i\mathcal E\Gamma^A\mathcal E+i\mathcal E^A\,\mathcal E\lambda+\frac{1}{3}\mathcal E^A\,\mathcal E^B\,\partial_B\,\phi
\end{eqnarray}
and the NS--NS three--form superfield strength is constrained as in \cite{Grassi:2009yj}
\begin{equation}
H=
-i\mathcal E^A\,\mathcal E\Gamma_A\Gamma_{11}\mathcal E
+i\mathcal E^B\mathcal E^A\,\mathcal E\Gamma_{AB}\Gamma_{11}\lambda
+\frac{1}{3!}\mathcal E^C\mathcal E^B\mathcal E^A\,H_{ABC}\,,
\label{eq:H}
\end{equation}
where $\lambda_\alpha(X,\Theta)$ is the dilatino superfield, $\phi(X,\Theta)$ is the dilaton and $\partial_A=\mathcal E_A{}^M\partial_M+\mathcal E_A{}^\mu\partial_\mu$. The dilatino superfield is not independent but is proportional to the spinor derivative of the dilaton \cite{Howe:2004ib}
\begin{equation}\label{lambda}
\lambda_\alpha=-\frac{i}{3}\,\partial_\alpha\,\phi\, :=-\frac{i}{3}\left(\mathcal E_\alpha{}^M\partial_M\phi+\mathcal E_\alpha{}^\mu\partial_\mu\phi\right)\,.
\end{equation}
The matrices $\mathcal E_A{}^M$, $\mathcal E_A{}^\mu$, etc.  in the definition of $\partial_A$ and $\partial_\alpha$ are the inverse supervielbeins.

We shall consider classical superstrings in  $AdS_4\times CP^3$ and $AdS_2\times S^2\times T^6$ superbackgrounds of type IIA supergravity. As we will see  many expressions turn out to be very similar which allows us to treat both cases simultaneously.
Nevertheless, strings in the $AdS_2\times S^2\times T^6$ superbackground of type IIB supergravity can also be treated in a similar fashion with only slight modifications of our formulas due to the same chirality of the $10d$ Majorana--Weyl spinors (see \cite{Sorokin:2011rr}).
\\{}\\
$\mathbf{AdS_4\times CP^3}$ is
parametrized by the $AdS_4$ coordinates $x^m$ $(m=0,1,2,3)$ and the $CP^3$ coordinates $y^{m'}$ $(m'=4,5,6,7,8,9)$. The vielbeins along $AdS_4$ are $e^a(x)=dx^m\,e_m{}^{a}(x)$  ($a=0,1,2,3$) and along $CP^3$ are $e^{a'}(y)=dy^{m'}e_{m'}{}^{a'}(y)$. The $10d$ vielbein is then $e^A(X^M)=(e^a(x),e^{a'}(y))$.

The $AdS_4$ curvature is
\begin{equation}
R_{ab}{}^{cd}=\frac{8}{R^2}\,\delta^c_{[a}\,\delta_{b]}^d \,,\qquad R^{ab}=-\frac{4}{R^2}\,e^a\,e^b\,,
\end{equation}
where $R$ is the $CP^3$ radius or twice the $AdS_4$ radius, and the $CP^3$ curvature is
\begin{equation}
R_{a'b'}{}^{c'd'}=-\frac{2}{R^2}\,(\delta^{c'}_{[a'}\,\delta_{b']}^{d'}+J_{[a'}{}^{c'}\,J_{b']}{}^{d'}+J_{a'b'}J^{c'd'})\,,
\end{equation}
where $J^{a'b'}$ is the K\"ahler form on $CP^3$.
\\{}\\
$\mathbf{AdS_2\times S^2\times T^6}$ is parametrized by the $AdS_2$ coordinates $x^m$ $(m=0,1)$, the $S^2$ coordinates $x^{\hat m}$ $(\hat m=2,3)$ and those of $T^6$ $y^{m'}$ $(m'=4,5,6,7,8,9)$. The corresponding  vielbeins are $e^{ a}=dx^{m}\,e_{m}{}^{a}(x)$  (${a}=0,1$),
$e^{\hat a}=dx^{\hat
m}\,e_{\hat m}{}^{\hat a}(\hat x)$ (${\hat a}=2,3$) and $e^{a'}(y)=dy^{a'}$. We will often combine the $AdS_2$ and $S^2$ indices into $\underline a=(a,\hat a)=0,1,2,3$.

The $AdS_2$ curvature is
\begin{equation}
R_{ab}{}^{cd}=\frac{8}{R^2}\,\delta^c_{[a}\,\delta_{b]}^d\,,\qquad R^{ab}=-\frac{4}{R^2}\,e^a\,e^b\,,
\end{equation}
where $R$ is \emph{twice} the $AdS_2$ (or $S^2$) radius, and the $S^2$ curvature is
\begin{equation}
R_{\hat a\hat b}{}^{\hat c\hat d}=-\frac{8}{R^2}\,\delta^{\hat c}_{[\hat a}\,\delta_{\hat b]}^{\hat d}\,,\qquad R^{\hat a\hat b}=\frac{4}{R^2}\,e^{\hat a}\,e^{\hat b}\,.
\end{equation}
The $10d$ curvature of $AdS_4\times CP^3$ and $AdS_2\times S^2\times T^6$ is denoted by $R_{AB}{}^{CD}$.

The $D=10$ gamma--matrices satisfy
\begin{equation}
\{\Gamma^A,\,\Gamma^B\}=2\eta^{AB}\,,\qquad \Gamma^A=(\Gamma^a,\,\Gamma^{a'})\,,\qquad a=0,1,2,3\qquad a'=4,\cdots,9\,.
\end{equation}
We also define
\begin{eqnarray}\label{gammas}
\gamma_5&=&i\Gamma^{0123},\nonumber\\
\gamma_7&=&i\Gamma^{456789},\nonumber\\
\Gamma_{11}&=&\gamma_5\gamma_7\,,
\end{eqnarray}
all of which square to one. The charge conjugation matrix is denoted $\mathcal C$. The matrices $\mathcal C$, $\mathcal C\Gamma_{\hat A\hat B\hat C}$ and $\mathcal C\Gamma_{\hat A\hat B\hat C\hat D}$ are anti-symmetric while $\mathcal C\Gamma_{\hat A}$, $\mathcal C\Gamma_{\hat A\hat B}$ and $\mathcal C\Gamma_{\hat A\hat B\hat C\hat D\hat E}$ are symmetric, where the indices are eleven dimensional, $\hat A=(A,11)$.

Finally we introduce a spinor projection matrix $\mathcal P_8$ which singles out an 8--dimensional subspace of the 32--dimensional space of spinors
\begin{equation}\label{P8}
\mathcal P_8=\frac{1}{8}(2-iJ_{a'b'}\Gamma^{a'b'}\gamma^7)\,,
\end{equation}
where $J_{a'b'}$ is the K\"ahler form on $CP^3$ or $T^6$. The complementary projection matrix which singles out a 24--dimensional subspace is then
\begin{equation}\label{P24}
\mathcal P_{24}=1-\mathcal P_8=\frac{1}{8}(6+iJ_{a'b'}\Gamma^{a'b'}\gamma^7)\,.
\end{equation}
Some useful identities satisfied by the gamma matrices and these projectors are given in Appendix \ref{sec:gammamatrices}.

In the $AdS_4\times CP^3$ case $\mathcal P_{24}$ singles out from $\Theta^\alpha$ 24 fermionic coordinates
$$\vartheta=\mathcal P_{24}\Theta$$
corresponding to the unbroken supersymmetries and, hence, to the Grassmann--odd directions of the supercoset $\frac{OSp(6|4)}{SO(1,3)\times U(3)}$, while the remaining eight
$$\upsilon=\mathcal P_8\Theta$$
are non--supercoset fermions.

In $AdS_2\times S^2\times T^6$ the role of the two projectors gets exchanged. $\mathcal P_{8}$ singles out from $\Theta^\alpha$ 8 fermionic coordinates
$$\vartheta=\mathcal P_{8}\Theta$$
corresponding to the unbroken supersymmetries and, hence, to the Grassmann--odd directions of the supercoset $\frac{PSU(1,1|2)\rtimes E(6)}{SO(1,1)\times U(1)\times SO(6)}$, while the remaining twenty four
$$\upsilon=\mathcal P_{24}\Theta$$
are non--supercoset fermions.

To treat the two cases simultaneously we shall always denote the coset fermions by $\vartheta$ and the non--coset ones by $\upsilon$.
The projector which singles out the \emph{coset} fermions will be denoted by $\mathcal P$, namely
\begin{equation}\label{P}
\vartheta=\mathcal P\Theta, \qquad\upsilon=(1-\mathcal P)\Theta.
\end{equation}

Let us note that as solutions of the type IIA supergravity equations of motion the $AdS_4\times CP^3$ and $AdS_2\times S^2\times T^6$ backgrounds also contain non--zero constant Ramond--Ramond $F_2$ and $F_4$ fluxes which are implicitly encoded in the form of the projectors \eqref{P8} and \eqref{P24}, namely \begin{equation}\label{slash}
\slashed F=-\frac{8i}{R}\mathcal P\gamma_\star,
\end{equation}
where
$$
\slashed F=e^\phi\left(-\frac{1}{2}\Gamma^{AB}\Gamma_{11}F_{AB}+\frac{1}{4!}\Gamma^{ABCD}F_{ABCD}\right)
$$
and $\gamma_\star$ stands for $\gamma_5$ in the $AdS_4\times CP^3$ case and for $\Gamma^{01}\gamma^7$ in the $AdS_2\times S^2\times T^6$ case. The explicit form of the RR fluxes can be found e.g. in \cite{Gomis:2008jt} and \cite{Sorokin:2011rr}\footnote{To be precise, eq. \eqref{slash} holds for the $AdS_2\times S^2\times T^6$ background with a non--zero $F_4$ and $F_2$ flux such that the latter has support on $S^2$ as in eq. (3.3) of \cite{Sorokin:2011rr}.}.

\subsection{Superstring action and equations of motion}
The Green--Schwarz superstring action in a general supergravity background \cite{Grisaru:1985fv}, written in terms of worldsheet differential forms, is
\begin{equation}
S=-\frac{T}{2}\int_\Sigma\,*\mathcal E^A\mathcal E^B\eta_{AB}+T\int_\Sigma\,B\,,
\end{equation}
where the pull--back to the worldsheet of the target--superspace quantities is understood, the star $*$ denotes the Hodge dual operation on the worldsheet and the wedge product of differential forms is implicit.
From this action one gets the superstring equations of motion which have the following form for our choice of the superspace constraints, eq. \eqref{TA} (see also \cite{Gomis:2008jt,Grassi:2009yj}). The fermionic field equations are
\begin{eqnarray}\label{Psi}
\Psi_\alpha\equiv i* {\mathcal E}^A\,(\Gamma_A {\mathcal E})_\alpha
-i {\mathcal E}^A\,(\Gamma_A\Gamma_{11} {\mathcal E})_\alpha
+\frac{i}{2}* {\mathcal E}^A {\mathcal E}_A\,\lambda_\alpha
+\frac{i}{2} {\mathcal E}^A {\mathcal E}^B\,(\Gamma_{AB}\Gamma_{11}\lambda)_\alpha=0\,,
\end{eqnarray}
and the bosonic field equations are
\begin{eqnarray}\label{B}
\mathcal B^A&\equiv&d*\mathcal E^A+*\mathcal E^B\Omega_B{}^A
+i* {\mathcal E}^A\, {\mathcal E}\lambda
+\frac{1}{3}(* {\mathcal E}^A {\mathcal E}^B\,  \partial_B\phi-* {\mathcal E}^B {\mathcal E}_B\,  \partial^A\phi)
\nonumber\\
&&{}
-i {\mathcal E}\Gamma^A\Gamma_{11} {\mathcal E}-2i {\mathcal E}^B\, {\mathcal E}\Gamma^A{}_B\Gamma_{11}\lambda
+\frac{1}{2} {\mathcal E}^C {\mathcal E}^B\,H^A{}_{BC}=0\,.
\end{eqnarray}

Note that in the $AdS_4\times CP^3$ case $\partial_A\phi=0$ \footnote{To check that in the $AdS_4\times CP^3$ case $\partial_A\phi=0$   one can use the fact that $\phi(\upsilon)$ and $\lambda_\alpha=-\frac{i}{3}\partial_\alpha\phi$ do not depend on $X^M$ and $\vartheta$ (see \cite{Gomis:2008jt}), so we have
$$
0=\partial_M\phi=\mathcal E_M{}^A\partial_A\phi+\mathcal E_M{}^\alpha\partial_\alpha\phi=\mathcal E_M{}^A\partial_A\phi+3i\mathcal E_M{}^{\alpha}\lambda_\alpha=0.
$$
If $\partial_A\phi=0$, from the above equation it follows that the contraction of the gravitino superfield $\mathcal E_M{}^{\alpha}$ with the dilatino $\lambda_\alpha$ of this supergravity solution is zero,
$
\mathcal E_M{}^{\alpha}\lambda_\alpha=0.
$
This can be checked using the explicit expressions for $\mathcal E^{\alpha}$ and $\lambda_\alpha$ derived in \cite{Gomis:2008jt}.}. We shall see that it is also zero in the $AdS_2\times S^2\times T^6$ case to the second order in $\upsilon$ and guess that it may also be true to all orders, because of the similarity between the two cases.

If we put the non--supercoset fermions $\upsilon$ to zero the equations of motion \eqref{Psi} and \eqref{B} reduce to those of the sigma--models on $\frac{OSp(6|4)}{SO(1,3)\times U(3)}$ and $\frac{PSU(1,1|2)\rtimes E(6)}{SO(1,1)\times U(1)\times SO(6)}$, respectively, with the equations of motion of the $T^6$ coordinates decoupled in the latter case. In our conventions the supercoset equations of motion have the following form
\begin{eqnarray}\label{Psicoset}
\Psi_{0\alpha}&=&i*E^A\,(\Gamma_AE)_\alpha-iE^A\,(\Gamma_A\Gamma_{11}E)_\alpha=0\,,\\
\mathcal B^A_0&=&\nabla*E^A-iE\Gamma^A\Gamma_{11}E=0\,,\label{Bcoset}
\end{eqnarray}
where $\nabla*E^A=d*E^A+*E_B\Omega_0^{BA}$ and $\Omega^{AB}_0(X,\vartheta)$ is the supercoset spin connection $\Omega_0^{AB}=\Omega^{AB}|_{\upsilon=0}$, and $E^A(X,\vartheta)=c^{-2}\mathcal E^A|_{\upsilon=0}$ and $E^\alpha(X,\vartheta)=c^{-1}\mathcal E^\alpha|_{\upsilon=0}$ (with $c$ being a constant dilaton factor) are the supercoset supervielbeins. They are the components of the Cartan form valued in the isometry supergroup $G$ (\emph{i.e.} $OSp(6|4)$ or $PSU(1,1|2)\rtimes E(6)$)
\begin{equation}\label{K}
K=g^{-1}dg(X,\vartheta)=\frac{1}{2}\Omega_0^{AB}M_{AB}+E^AP_A+Q_\alpha E^\alpha\,, \qquad g(X,\vartheta) \in G/H\,.
\end{equation}
The algebra of the isometry generators and the explicit form of the supercoset bosonic and fermionic supervielbeins, and the spin connection are given in Appendix A. The isometry algebra is invariant under the following action of the $\mathbb Z_4$--automorphism on the generators $T=(M,P,Q)$
\begin{equation}\label{Z4auto}
\Omega(T)\equiv \Omega^{-1}T\Omega\,\qquad \Omega (M_{AB})=M_{AB},
 \qquad
 \Omega (P_{A})=-P_{A},
 \qquad
 \Omega (Q)=-iQ\Gamma _{11}, \qquad \Omega^4=1\,.
\end{equation}
Note that in the $AdS_2\times S^2\times T^6$ case the $T^6$ translation generators $P_{a'}$ also have $\mathbb Z_4$--grading one, as those of $AdS_2\times S^2$.

When the non--coset fermions are non--zero, the supercoset field equations \eqref{Psicoset} and \eqref{Bcoset} acquire non--zero right--hand sides
\begin{equation}\label{PsiB}
\Psi_{0}=\mathcal O(\upsilon)\,,\qquad \mathcal B_0=\mathcal O(\upsilon)
\end{equation}
which should be taken into account when extending the supercoset Lax connection to a zero--curvature Lax connection of the complete theory (see Section 3).

\subsection{Supercoset Lax connection}
The equations of motion of the superstring on the semi--symmetric supercoset spaces with $\mathbb Z_4$--grading \eqref{K} and \eqref{Z4auto} admit a Lax representation which implies classical integrability of the corresponding sigma--model \cite{Bena:2003wd,Adam:2007ws,Arutyunov:2008if,Stefanski:2008ik}. This means that from the components of the Cartan form \eqref{K} pulled--back on the worldsheet and their $2d$ Hodge duals one can construct a Lax connection (depending on a spectral parameter) which has zero curvature provided that the equations of motion \eqref{Psicoset} and \eqref{Bcoset} are satisfied. And vice versa the zero curvature of the Lax connection implies the field equations. In our notation and conventions the supercoset Lax connection has the following form
\begin{equation}\label{Lcoset}
L_{coset}=\frac{1}{2}\Omega_0^{AB}M_{AB}+(1+\alpha_1)E^AP_A+\alpha_2\ast E^AP_A+Q(\beta_2+\beta_1\Gamma_{11})E\,,
\end{equation}
where $\alpha_1$, $\alpha_2$, $\beta_1$ and $\beta_2$ are numerical parameters whose values are determined by requiring the zero--curvature condition
\begin{equation}\label{R0}
dL_{coset}-L_{coset}\wedge L_{coset}=0
\end{equation}
to hold on the mass--shell \eqref{Psicoset} and \eqref{Bcoset}. This gives the following relations between the parameters
\begin{equation}\label{alpha}
\alpha_2^2=2\alpha_1+\alpha_1^2\, 
\end{equation}
and
\begin{eqnarray}\label{beta}
\beta_1=\mp\sqrt{\frac{\alpha_1}{2}}\,,
\qquad \beta_2=\pm\frac{\alpha_2}{\sqrt{2\alpha_1}}
\,.
\end{eqnarray}
They can therefore be expressed in terms of a single spectral parameter $\tt x$ as follows
\begin{equation}\label{x}
\alpha_1=\frac{2{\tt x}^2}{1-{\tt x}^2}, \qquad \alpha_2=\frac{2{\tt x}}{1-{\tt x}^2}\,\qquad \beta_1=-\frac{i{\tt x}}{\sqrt{{\tt x}^2-1}}\,,\qquad \beta_2=\frac{i}{\sqrt{{\tt x}^2-1}}\,.
\end{equation}
It is very useful for further analysis to specify the properties of the $32\times32$ matrix
\begin{equation}\label{eq:V}
V=\beta_2+\beta_1\Gamma_{11}\,,
\end{equation}
which enters the Lax connection \eqref{Lcoset}. It is easily seen to satisfy the relations
\begin{equation}\label{G}
V^2=1+\alpha_1-\alpha_2\Gamma_{11}\,, \quad {VV}^\dagger=\beta_2^2-\beta_1^2=1\,,\quad (V^\dagger)^\alpha{}_\beta=-(\mathcal {C}V^\mathrm{T}\mathcal C)^\alpha{}_\beta=(\beta_2-\beta_1\Gamma_{11})^\alpha{}_\beta\,,
\end{equation}
where $\mathcal C$ denotes the anti-symmetric charge-conjugation matrix (see Section \ref{sec:notation}). Therefore $V\in Sp(32)$.

It is easy to check that the Lax connection \eqref{Lcoset} is invariant under the $\mathbb Z_4$--transformations of the generators  \eqref{Z4auto} accompanied by the inversion of the spectral parameter
\begin{equation}\label{Ox}
\Omega(\tt x) = \frac{1}{\tt x}\,,
\end{equation}
which implies that
\begin{equation}\label{Oalpha}
\alpha_1\rightarrow-\alpha_1-2\,,\qquad \alpha_2\rightarrow -\alpha_2\,,\qquad V\rightarrow i\Gamma_{11}V\,.
\end{equation}
Namely,
$$
\Omega(L_{coset}({\tt x}))=\Omega^{-1}L_{coset}(\frac{1}{\tt x})\,\Omega=L_{coset}({\tt x})\,.
$$

Note that in the $AdS_2\times S^2\times T^6$ case the first term of \eqref{Lcoset} may in general include the $SO(6)$ spin connection on $T^6$ whose curvature is zero. Therefore, it can be gauged away by performing a suitable gauge transformation of $L_{coset}$, and the resulting Lax connection will contain only terms associated with the  $U(1)^6$ `translations'  $P_{a'}$ along $T^6$ which completely decouple from the $PSU(1,1|2)$ part and, therefore, can be taken with arbitrary coefficients.

The explicit dependence of the supercoset Lax curvature on the left--hand sides of the supercoset field equations  \eqref{Psicoset} and \eqref{Bcoset} looks as follows
\begin{equation}\label{R01}
dL_{coset}-L_{coset}\wedge L_{coset}=\alpha_2 (\mathcal B_0^AP_A-\frac{1}{R}QV^\dagger\gamma_\star \Psi_0)\,,
\end{equation}
where again $\gamma_\star $ stands for $\gamma^5$ in the $AdS_4\times CP^3$ case and for $\Gamma^{01}\gamma^7$ in the $AdS_2\times S^2\times T^6$ case (see eqs. \eqref{gammas} and \eqref{gammasA}).

\section{The Lax connection of the complete GS superstring to quadratic order in non--coset fermions $\upsilon$}\label{comleteL}
We are now ready to extend the supercoset Lax connection \eqref{Lcoset} with terms that include contributions from the string fermionic modes $\upsilon$ associated with broken target--space supersymmetries
\begin{equation}
L=L_{coset}(X,\vartheta)+\alpha_2L'(X,\vartheta,\upsilon)\,,
\label{eq:Lax-connection}
\end{equation}
where the factor of $\alpha_2$ in front of $L'$ is due to the same factor on the right--hand side of \eqref{R01}. The correction $L'$ which is aimed at canceling the r.h.s. of \eqref{R01}, turns out to take the same form for both the $AdS_4\times CP^3$ and $AdS_2\times S^2\times T^6$ case and looks as follows
\begin{eqnarray}\label{L'}
L'&=&
-\frac{i}{R}Q\gamma_\star \left[
*(E^A+2i\upsilon\Gamma^AE)\,\Gamma_A V \upsilon
-(E^A+2i\upsilon\Gamma^AE)\,\Gamma_A\Gamma_{11} V \upsilon\right]\nonumber\\
&&
-\frac{i}{R}Q\gamma_\star \left[i(\upsilon\Gamma^A\Gamma_{11}E)\,\Gamma_A V \upsilon
+i(\upsilon\Gamma^AE)\,\Gamma_A\Gamma_{11} V \upsilon
\right]
\nonumber\\
&&{}
+(
2i\upsilon\Gamma^A*E+i\upsilon\Gamma^A*\nabla\upsilon-\frac{2}{R}*E^B\,\upsilon\Gamma^A\mathcal P\gamma_\star \Gamma_B\upsilon
)\,P_A
\nonumber\\
&&{}
+(
2i\upsilon\Gamma^A\Gamma_{11}E+i\upsilon\Gamma^A\Gamma_{11}\nabla\upsilon-\frac{2}{R}E^B\,\upsilon\Gamma^A\Gamma_{11}\mathcal P\gamma_\star \Gamma_B\upsilon
)\,P_A
\nonumber\\
&&{}
+\frac{i}{8}(*E^C\,\upsilon\Gamma_C{}^{DE} V ^2\upsilon-E^C\,\upsilon\Gamma_C{}^{DE}\Gamma_{11} V ^2\upsilon)\,R_{DE}{}^{AB}\,M_{AB}
\,,
\end{eqnarray}
where the matrix $V$ has been introduced in \eqref{eq:V}, and $\gamma_\star $, $\mathcal P$  and $R_{DE}{}^{AB}$ are defined for $AdS_4\times CP^3$ and $AdS_2\times S^2\times T^6$ in Section \ref{sec:notation} and Appendix \ref{sec:supercosets}. The covariant derivative is defined with respect to the  supercoset spin connection $\Omega_0^{AB}=\Omega^{AB}|_{\upsilon=0}$, i.e. $\nabla \upsilon=(d-\frac{1}{4}\Omega_0^{AB}\Gamma_{AB})\upsilon$.

To construct $L'(X,\vartheta,\upsilon)$ one had to know the form of the right--hand sides of eqs. \eqref{Psicoset} and \eqref{Bcoset}. For the $AdS_4\times CP^3$ case these can be given to all orders in $\upsilon$ since the explicit form of the geometry of the $AdS_4\times CP^3$ superspace is known \cite{Gomis:2008jt}. Because of technical complications, we have however restricted the construction of the Lax connection to the second order in $\upsilon$ only.

In the $AdS_2\times S^2\times T^6$ case the explicit form of the geometry of the complete target superspace is not know, so our strategy was somewhat opposite to that of the $AdS_4\times CP^3$ case. We have assumed that in the $AdS_2\times S^2\times T^6$ case the Lax connection has a similar form to the $AdS_4\times CP^3$ Lax connection, i.e. eq. \eqref{L'} with the appropriate replacement of the supersymmetry projector $\mathcal P$ and the product of gamma--matrices $\gamma_\star $ appearing in the definition of the isometry superalgebra (see Appendix \ref{sec:supercosets}). Then, requiring that the $AdS_2\times S^2\times T^6$ Lax connection has zero curvature we have reconstructed the superstring equations and the form of the $AdS_2\times S^2\times T^6$ superbackground up to the second order in the non--coset fermions and checked that it indeed satisfies the constraints of type IIA supergravity.

\subsection{$AdS_4\times CP^3$ case}\label{AdS4}
Taking the expressions for the quantities defining the full $AdS_4\times CP^3$ supergeometry \cite{Gomis:2008jt} and expanding them to the second order in $\upsilon$ we get
\begin{eqnarray}\label{geometry}
\mathcal E^A&=&c^2(1-\frac{1}{R}\upsilon\gamma^5\upsilon)(E^A+2i\upsilon\Gamma^AE+i\upsilon\Gamma^AD\upsilon)+\mathcal O(\upsilon^3)\,,\nonumber\\
\mathcal P\mathcal E&=&c(1-\frac{1}{2R}\upsilon\Gamma^b\gamma^5\upsilon\,\Gamma_b+\frac{1}{R}\upsilon\Gamma^b\gamma^7\upsilon\,\Gamma_b\Gamma_{11}+\frac{1}{2R}\upsilon\upsilon\,\gamma_5)E+\mathcal O(\upsilon^3)\,,\nonumber\\
(1-\mathcal P)\mathcal E&=&cD\upsilon+\mathcal O(\upsilon^3)\nonumber\,,\\
\lambda&=&\frac{2i}{cR}\gamma^5\upsilon+\mathcal O(\upsilon^3)\nonumber\,,\\
H_{abc}&=&-\frac{12i}{c^2R^2}\upsilon\Gamma_{abc}\Gamma_{11}\upsilon+\mathcal O(\upsilon^3)\,,
\end{eqnarray}
where $c=e^{\frac{1}{6}\phi_0}=\left(\frac{R}{kl_p}\right)^{1/4}$, $\phi_0$ is the value of the dilaton for the $AdS_4\times CP^3$ supergravity solution, $k$ is the `Chern--Simons' level, $l_p$ is the $11d$ Planck length and
\begin{equation}\label{DV1}
D\upsilon=(\nabla+\frac{i}{R}E^a\,\gamma_5\Gamma_a)\upsilon\,.
\end{equation}
Remember that in the above expressions $E^A(X,\vartheta)$, $E^\alpha(X,\vartheta)$ and $\Omega_0^{AB}(X,\vartheta)$ are the components of the Cartan form of the supercoset $\frac{OSp(6|4)}{SO(1,3)\times U(3)}$  (see Section \ref{sec:notation} and Appendix \ref{SG}). Note also that though in the background under consideration the purely bosonic part of the NS-NS flux $H_{ABC}(X)$ is zero, its superfield extension is non--trivial and depends on the non--coset fermionic coordinates.

We can now insert the expressions \eqref{geometry} into the complete equations of motion \eqref{Psi} and \eqref{B} and thus find the corrections to the supercoset equations \eqref{PsiB}. For this we should also know the form of the spin connection $\Omega^{AB}(X,\vartheta,\upsilon)$ which was not derived in \cite{Gomis:2008jt}. The expression for $\Omega^{AB}(X,\vartheta,\upsilon)$ can be obtained by analyzing the torsion constraint \eqref{TA} and has the following form to the second order in $\upsilon$
\begin{eqnarray}\label{Omega}
\Omega^{AB}&=&\Omega_0^{AB}+\frac{2}{R}\Big(
-\delta^A_{a'}\delta^B_{b'}\upsilon\Gamma^{a'b'}\gamma_5E
+\delta^A_a\delta^B_b\,\upsilon\Gamma^{ab}\gamma_5D\upsilon
+\frac{i}{R}\delta^A_{a'}\delta^B_{b'}\,E^c\,\upsilon\Gamma^{a'b'}{}_c\upsilon
\nonumber\\
&&{}
-\frac{2i}{R}\delta^{[A}_{a'}\delta^{B]}_b\,E^{c'}\,\upsilon\Gamma^{a'b}{}_{c'}\upsilon
\Big)+\mathcal O(\upsilon^3)\,.
\end{eqnarray}
Notice that (due to the last term in \eqref{Omega}) the spin connection takes values in the whole $D=10$ Lorentz algebra $so(1,9)$ rather than in the stability subalgebra $so(1,3)\oplus u(3)$ of the bosonic coset $AdS_4\times CP^3$. This reflects the fact that the complete $AdS_4\times CP^3$ superspace is not a supercoset.

The explicit form of the corrections to the equations of motion  \eqref{Psicoset} and \eqref{Bcoset} are given in Appendix \ref{eom}. Thus, the correction $L'$ to the supercoset Lax connection which cancels the contributions of the right--hand sides of these equations to the Lax curvature \eqref{R01} has been found to be eq. \eqref{L'}.

\subsection{$AdS_2\times S^2\times T^6$ Lax connection and supergeometry}\label{AdS2}
We now pass to the consideration of the $AdS_2\times S^2\times T^6$ case. As we have already mentioned, a problem that we meet is that here the explicit form of the corresponding $10d$ superspace with 32 fermionic directions is unknown. What is known is the supercoset structure of its sub--superspace obtained by putting to zero 24 fermionic coordinates $\upsilon$ and, on the other hand, the structure of the complete $AdS_2\times S^2\times T^6$ superspace to the second order in the 32 fermions $\Theta$  (see \cite{Sorokin:2011rr} for more details and references). Our goal is to take a step further and to find the explicit form of the $AdS_2\times S^2\times T^6$ superbackground to all orders in the 8 coset fermions $\vartheta$ and to quadratic order in 24 non--coset fermions $\upsilon$. To this end, we assume that the $AdS_2\times S^2\times T^6$  Lax connection has the form of \eqref{eq:Lax-connection} and \eqref{L'} in which now $L_{coset}$ and $L'$ are constructed with the use of the Cartan forms, the curvature of $\frac{PSU(1,1|2)\rtimes E(6)}{SO(1,1)\times U(1)\times SO(6)}$ and the structure of its isometry superalgebra given in Appendix \ref{sec:supercosets}. Namely, in \eqref{L'} we now take $\mathcal P$ to be the supersymmetry projector $\mathcal P_8$ and $\gamma_\star =\Gamma^{01}\gamma^7$ as in the superalgebra of $PSU(1,1|2)\rtimes E(6)$.

Then, requiring that the curvature of the Lax connection \eqref{eq:Lax-connection} and \eqref{L'} vanishes, we get the form of the superstring equations \eqref{Psi} and \eqref{B} in $AdS_2\times S^2\times T^6$ as a deformation of the $\frac{PSU(1,1|2)\rtimes E(6)}{SO(1,1)\times U(1)\times SO(6)}$ sigma--model field equations to the second order in $\upsilon$ and consequently reconstruct to the same order the form of the supervielbeins, the spin connection and the NS--NS superfield strength of the complete $AdS_2\times S^2\times T^6$ superbackground.

Using the expressions given in Appendix \ref{sec:dL} the curvature of the Lax connection \eqref{eq:Lax-connection} with $L'$ given in eq. \eqref{L'} and valued in the $PSU(1,1|2)\rtimes E(6)$ superalgebra of Appendix \ref{sec:supercosets} can be assembled to have the following form
\begin{eqnarray}
dL-LL&=&
\frac{\alpha_2}{c^3}\Big[
\frac{1}{4}(
\upsilon(V^\dagger)^2\Gamma^{\underline{cd}}\Psi
+\frac{ic}{2}\mathcal B^{\underline a}\,\upsilon\Gamma_{\underline a}{}^{\underline{cd}}V^2\upsilon
)\,R_{\underline{cd}}{}^{\underline{ef}}M_{\underline{ef}}
+c\mathcal B^A\,P_A
\nonumber\\
&&{}
+\frac{2i}{R}\upsilon\gamma_\star \Gamma^A(1-\mathcal P)\Psi\,P_A
+\frac{c}{R}\mathcal B^B\,\upsilon\Gamma^A{}_B\gamma_\star \upsilon\,P_A
+\frac{c}{R}\mathcal B^B\,\upsilon\Gamma^A\mathcal P\Gamma_B\gamma_\star \upsilon\,P_A
\nonumber\\
&&{}
-\frac{1}{R}QV^\dagger\gamma_\star \Psi
-\frac{ic}{R}\mathcal B^A\,QV^\dagger\gamma_\star \Gamma_A\upsilon
-\frac{1}{4R^2}QV^\dagger\gamma_\star \Gamma_{BC}\Psi\,\upsilon\Gamma^{BC}\gamma_\star \upsilon
\nonumber\\
&&{}
-\frac{1}{2R^2}QV^\dagger\gamma_\star \gamma_7\Gamma_B\Psi\,\upsilon\Gamma^B\gamma_\star \gamma_7\upsilon
-\frac{1}{2R^2}QV^\dagger\gamma_\star \Gamma_B\Gamma_{11}\Psi\,\upsilon\Gamma^B\gamma_\star \Gamma_{11}\upsilon
\nonumber\\
&&{}
-\frac{1}{2R^2}QV^\dagger\gamma_\star \Gamma_{\underline ab'}\gamma_5\gamma_\star \Psi\,\upsilon\Gamma^{\underline ab'}\gamma_5\upsilon
-\frac{1}{4R^2}QV^\dagger\gamma_\star \Gamma_{\underline{ab}c'}\gamma_\star \Psi\,\upsilon\Gamma^{\underline{ab}c'}\upsilon
\nonumber\\
&&{}
+\frac{\alpha_2}{2R^2}\Big(
\upsilon\Gamma^{\underline a}\gamma^5\upsilon\,QV\gamma^7\gamma_\star \Gamma_{\underline a}\gamma_\star \Psi
+\upsilon\Gamma^{\underline a}\gamma^7\upsilon\,QV\gamma^5\gamma_\star \Gamma_{\underline a}\gamma_\star \Psi
\nonumber\\
&&{}
+2\upsilon\Gamma^{a'}\gamma_\star \Gamma_{11}\upsilon\,QV\Gamma_{a'}\gamma_\star \Psi
+2\upsilon\Gamma^{a'}\gamma_\star \upsilon\,QV\Gamma^{11}\Gamma_{a'}\gamma_\star \Psi
\Big)
\Big]+\mathcal O(\upsilon^3)\,,
\label{eq:dLAdS2}
\end{eqnarray}
where, if the curvature is zero, $\Psi_\alpha$ and $\mathcal B^A$ should vanish and hence should coincide with the equations of motion of the  $AdS_2\times S^2\times T^6$ superstring.

Comparing the form of these  $\Psi_\alpha$ and $\mathcal B^A$ with the Green--Schwarz superstring equations \eqref{Psi} and \eqref{B} we find that the $AdS_2\times S^2\times T^6$ supergeometry is described by the following supervielbeins, dilatino and spin connection
\begin{eqnarray}
\mathcal E^A&=&c^2(1+\frac{1}{R}\upsilon\gamma_\star \upsilon)(E^A+2i\upsilon\Gamma^AE+i\upsilon\Gamma^AD\upsilon)
+\mathcal O(\upsilon^3)\nonumber\\
\mathcal P\mathcal E&=&
c\mathcal P(1
+\frac{1}{2R}\upsilon\Gamma^B\gamma_\star \gamma_7\upsilon\,\Gamma_B\gamma_7-\frac{1}{2R}\upsilon\Gamma^B\gamma_\star \Gamma_{11}\upsilon\,\Gamma_B\Gamma_{11}-\frac{1}{4R}\upsilon\Gamma^{BC}\gamma_\star \upsilon\,\Gamma_{BC})E
+\mathcal O(\upsilon^3)\nonumber\\
(1-\mathcal P)\mathcal E&=&c[D\upsilon
+(1-\mathcal P)(\frac{1}{2R}\upsilon\Gamma^B\gamma_\star \gamma_7\upsilon\,\Gamma_B\gamma_7
-\frac{1}{2R}\upsilon\Gamma^B\gamma_\star \Gamma_{11}\upsilon\,\Gamma_B\Gamma_{11}
-\frac{1}{4R}\upsilon\Gamma^{BC}\gamma_\star \upsilon\,\Gamma_{BC}
\nonumber\\
&&{}
-\frac{1}{2R}\upsilon\Gamma^{\underline bc'}\Gamma_{11}\upsilon\,\gamma_\star \Gamma_{\underline bc'}\Gamma_{11}
-\frac{1}{4R}\upsilon\Gamma^{\underline{bc}d'}\upsilon\,\gamma_\star \Gamma_{\underline{bc}d'}
)E]
+\mathcal O(\upsilon^3)\nonumber\\
\lambda&=&-\frac{2i}{cR}\gamma_\star \upsilon+\mathcal O(\upsilon^3)\,,
\label{eq:AdS2vielbeins}
\end{eqnarray}
\begin{eqnarray}
\Omega^{AB}&=&\Omega_0^{AB}
-\frac{2}{R}\delta^{[A}_{a'}\delta^{B]}_{b'}\upsilon\gamma_\star \Gamma^{a'b'}E
-\frac{1}{R}\upsilon\Gamma^{[A}(1-\mathcal P)\Gamma^{B]}\gamma_\star D\upsilon
-\frac{1}{R}\upsilon\gamma_\star \Gamma^{[A}(1-\mathcal P)\Gamma^{B]}D\upsilon
\nonumber\\
&&{}
+\frac{i}{R^2}\delta^{[A}_{a'}\delta^{B]}_{b'}E^C\,\upsilon\Gamma^{a'}(1-\mathcal P)\Gamma^{b'}\gamma_\star \Gamma_C\gamma_\star \upsilon
+\frac{i}{R^2}\delta^{[A}_{a'}\delta^{B]}_{b'}E^{c'}\,\upsilon\Gamma^{a'}\mathcal P\Gamma^{b'c'}\upsilon
\nonumber\\
&&{}
-\frac{2i}{R^2}\delta^{[A}_{a'}\delta^{B]}_{\underline b}E^{c'}\,\upsilon\Gamma^{a'}(1-\mathcal P)\Gamma^{\underline b}\Gamma_{c'}\upsilon
+\frac{2i}{R^2}\delta^{[A}_{a'}\delta^{B]}_bE^{\hat c}\,\upsilon\Gamma^{a'b}{}_{\hat c}\upsilon
\nonumber\\
&&{}
-\frac{2i}{R^2}\delta^{[A}_{a'}\delta^{B]}_{\hat b}E^c\,\upsilon\Gamma^{a'\hat b}{}_c\upsilon
-\frac{2i}{R^2}\delta^{[A}_a\delta^{B]}_{\underline b}E^{c'}\,\upsilon\Gamma^{a\underline b}{}_{c'}\upsilon
+\mathcal O(\upsilon^3)\,,
\label{eq:AdS2Omega}
\end{eqnarray}
where
\begin{eqnarray}\label{Dv}
D\upsilon=(\nabla+\frac{i}{R}E^B\,(1-\mathcal P)\Gamma_B\gamma_\star )\upsilon\,
\end{eqnarray}
and the constant $c=e^{\frac{1}{6}\phi_0}$, where $\phi_0$ is the dilaton vacuum expectation value of the $AdS_2\times S^2\times T^6$ supergravity solution.

Finally, the NS--NS three--form field strength turns out to be
\begin{eqnarray}
H_{ABC}&=&
\frac{6i}{c^2R^2}\Big(
\upsilon\gamma_\star \Gamma_{ABC}\Gamma_{11}\gamma_\star \upsilon
-\delta_{[A}^{\underline a}\,\upsilon\gamma_\star \Gamma_B\mathcal P\Gamma_{C]}\Gamma_{\underline a}\gamma_\star \Gamma_{11}\upsilon
-\delta_{[A}^d\delta_B^e\,\upsilon\Gamma_{C]de}\Gamma_{11}\upsilon
\nonumber\\
&&{}
+\delta_{[A}^{\hat d}\delta_B^{\hat e}\,\upsilon\Gamma_{C]\hat d\hat e}\Gamma_{11}\upsilon
\Big)
+\mathcal O(\upsilon^3)\,.
\label{eq:AdS2HABC}
\end{eqnarray}

It remains to verify that this is indeed the correct form of the supergeometry, i.e. that the above expressions solve the supergravity constraints to the relevant order. This is not guaranteed since they were derived from the flatness of a Lax connection which we simply postulated. It is known that it is enough to demonstrate that the Green-Schwarz string action possesses kappa-symmetry in order to say that the background is a solution to the supergravity equations. This in turn simply amounts to verifying that the components of the torsion $T^A$ and NS--NS superfield strength $H$ have the appropriate components involving fermionic supervielbeins whose form is dictated by the supergravity constraints \eqref{TA} and \eqref{eq:H}.

Indeed, with the above choice of the spin connection \eqref{eq:AdS2Omega} it is not hard to verify that the type IIA supergravity torsion constraint \eqref{TA} is satisfied to the quadratic order in $\upsilon$. As far as the constraint \eqref{eq:H} is concerned we should substitute into it the expressions \eqref{eq:AdS2vielbeins} and \eqref{eq:AdS2HABC} and check that the resulting superform $H$ is closed. This can indeed be verified but the calculation is somewhat lengthy and we leave the details for Appendix \ref{sec:dH}.

Thus we have constructed the supergeometry of the $AdS_2\times S^2\times T^6$ solution of type IIA supergravity to all orders in the 8 supercoset fermions $\vartheta$ and to the second order in the 24 non--supercoset fermions $\upsilon$. Although in this paper we have not obtained the RR superfield strengths $F_2$ and $F_4$, which are needed for the construction of D--brane actions in this superbackground,  these can be found using the Bianchi identities, the corresponding superspace constraints given in \cite{Grassi:2009yj} and the above form of the supervielbeins.

By constructing the zero--curvature Lax connection we have demonstrated that the GS string in $AdS_2\times S^2\times T^6$ is classically integrable up to quadratic order in the non--supercoset fermions and to all orders in the coset fermions thus extending the results of \cite{Sorokin:2011rr} which considered all the fermions to the second order only.

\section{Properties of the Lax connection}
\subsection{$\mathbb Z_4$--invariance}
The Lax connection \eqref{eq:Lax-connection} with $L_{coset}$ and $L'$ given, respectively, in \eqref{Lcoset} and \eqref{L'} is invariant under the $\mathbb Z_4$--transformations \eqref{Z4auto} of the isometry generators and the inversion of the spectral parameter \eqref{Ox} and \eqref{Oalpha}. This demonstrates that the contribution of the non--coset fermions does not spoil the $\mathbb Z_4$--symmetry of the \emph{supercoset} Lax connection which is of crucial importance for the derivation of the algebraic curve and the Bethe ansatz equations, both classical and quantum \cite{Beisert:2005bm,Gromov:2008bz,Babichenko:2009dk,SchaferNameki:2010jy}.  The $\mathbb Z_4$--invariance induces the corresponding conjugation symmetry of the monodromy matrix of the Lax connection
$$\Omega^{-1}\, M(1/{\tt x})\, \Omega = M({\tt x})$$
used for the construction of the algebraic curve \footnote{We thank Kostya Zarembo for the discussion of these points.}.

\subsection{The Lax connection and conserved currents}
Let us now present an interesting observation how the Lax connection $L=L_{coset}+\alpha_2L'$ can be related to the conserved current associated with the superisometries. First of all notice that in the limit  $\alpha_2=\epsilon\rightarrow0$, in which
\begin{equation}\label{eps}
\alpha_1=\frac{1}{2}\epsilon^2+\mathcal O(\epsilon^4)\,,\qquad\beta_1=-\frac{1}{2}\epsilon+\mathcal O(\epsilon^2)\,, \qquad \qquad \beta_2=1+\mathcal O(\epsilon^2)\,,
\end{equation}
and
$$
V=\beta_2+\beta_1\Gamma_{11}\rightarrow 1, \qquad V^\dagger=\beta_2-\beta_1\Gamma_{11}\rightarrow 1\,,
$$
the Lax connection reduces to
\begin{equation}\label{Leps}
L=K+\mathcal O(\epsilon)\,,
\end{equation}
where $K(X,\vartheta)$ is the supercoset Cartan form introduced in \eqref{K}.

In fact, the term denoted by $\mathcal O(\epsilon)$ in the (gauge--transformed) Lax connection is the worldsheet Hodge dual of a superstring conserved current $J$ associated with the background superisometries, namely
\begin{equation}\label{J}
*J=\lim_{\epsilon\rightarrow0}\,\frac{1}{\epsilon}\,(gLg^{-1}-dgg^{-1})= g\lim_{\epsilon\rightarrow0}\frac{L-K}{\epsilon}g^{-1}\,,
\end{equation}
where $g(X,\vartheta)$ is the superisometry group element determining $K$ in \eqref{K}.

The conservation of $J$, i.e. $d*J=0$, follows from the flatness of the Lax connection and the Cartan form
\begin{equation}
dL-LL=0\,,\qquad dK-KK=0\,.
\end{equation}
Indeed, in view of \eqref{Leps}, we have
\begin{eqnarray}
d*J&=&
g\lim_{\epsilon\rightarrow0}\frac{dL-dK}{\epsilon}g^{-1}
-*JgKg^{-1}
-gKg^{-1}*J
\nonumber\\
&=&
g\lim_{\epsilon\rightarrow0}\frac{(L-K)K+K(L-K)}{\epsilon}g^{-1}
-*JgKg^{-1}
-gKg^{-1}*J
=0\,.
\end{eqnarray}
Note that in the case of the supercoset Lax connection \eqref{Lcoset} (when $\upsilon=0$), the current constructed in this way coincides with the conserved current of the $G/H$ sigma--model considered in \cite{Bena:2003wd,Arutyunov:2008if,Sorokin:2010wn}
\begin{equation}\label{Jcoset}
J_{coset}=g(E^AP_A-\frac{1}{2}Q\Gamma_{11}*E)g^{-1}\,.
\end{equation}

We can now write the correction \eqref{L'} to the Lax connection in terms of (transformed) components of the conserved current as
\begin{equation}
L'=g^{-1}*(\tilde J-\tilde J_{coset})g\,,
\end{equation}
where $\tilde J$ and $\tilde J_{coset}$ are, respectively, the complete conserved current (to second order in $\upsilon$) \eqref{J} and the conserved current of the supercoset model \eqref{Jcoset}, and the tilde means that in their expressions we perform the following substitutions of spinorial quantities
$$E\rightarrow V^\dagger E,\qquad  \nabla\upsilon\rightarrow V^\dagger\nabla\upsilon \qquad {\rm and} \qquad\upsilon\rightarrow V\upsilon.$$
For instance,
\begin{equation}
\tilde J_{coset}=g(E^AP_A-\frac{1}{2}Q\Gamma_{11}V^\dagger*E)g^{-1}\,.
\end{equation}

Whether this fact is of some deeper significance remains to be understood. Perhaps, a better understanding of this could lead to a proposal for the complete Lax connection to all orders in the non--coset fermions. We leave this problem for future analysis.

\subsection{Relation to the Lax connections constructed in \cite{Sorokin:2010wn,Sorokin:2011rr}}
In \cite{Sorokin:2010wn,Sorokin:2011rr} Lax connections for the superstring in $AdS_4\times CP^3$ and $AdS_2\times S^2\times T^6$ have been constructed (up to second order in fermions) using components of the Noether currents of the corresponding superisometries $OSp(6|4)$ and $PSU(1,1|2)\times U(1)^6$ (note that only the Abelian $U(1)^6$--currents of the $E(6)$ isometries enter the Lax connection)
\begin{equation}\label{Jcurrent}
J=J_{\mathcal B}+J_{susy}\,,
\end{equation}
where the conserved current $J_{\mathcal B}$ of the bosonic isometries has the form
\begin{equation}\label{JB}
J_{\mathcal B}=J_1+J_2\,,
\end{equation}
\begin{eqnarray}\label{J12}
J_1&=&\Big(e^A(X)+i\Theta\Gamma^A {\nabla}\Theta+i\Theta\Gamma^A\Gamma_{11}*\nabla\Theta\nonumber\\
&&-\frac{2}{R}e^B\,\Theta\Gamma^A\mathcal P\gamma_\star \Gamma_B\Theta
-\frac{2}{R}*e^B\,\Theta\Gamma^A\mathcal P\gamma_\star\Gamma_{11} \Gamma_B\Theta\,\Big)(gP_A\,g^{-1})|_{\vartheta=0},\nonumber\\
\\
J_2&=&\frac{i}{8}\left(e^C\,\Theta\Gamma^{AB}{}_C\Theta
-\ast e^C\,\Theta\Gamma^{AB}{}_C\Gamma_{11}\Theta\right)R_{AB}{}^{DE}\,(g\,M_{DE}\,g^{-1})|_{\vartheta=0}\hspace{50pt}\nonumber
\end{eqnarray}
and the supersymmetry current $J_{susy}$ is
\begin{equation}\label{Jsusy}
J_{susy}=\frac{i}{R}\,(g Q g^{-1})|_{\vartheta=0}\,\gamma_\star\,(\ast
e^A\,\Gamma_A\Gamma_{11}\Theta-e^A\,\Gamma_A\Theta  )\,.
\end{equation}
In \eqref{J12} and \eqref{Jsusy} the isometry group element \eqref{K} is evaluated at $\vartheta=0$ \footnote{Note that $K_A(X)=(gP_A\,g^{-1})|_{\vartheta=0}$ and $\Xi(X)=\gamma_\star(gQ g^{-1})|_{\vartheta=0}$ are simply the Killing vector and the Killing spinor of the superisometries (see  \cite{Sorokin:2010wn,Sorokin:2011rr} for more details).}.

The Lax connection constructed in \cite{Sorokin:2010wn,Sorokin:2011rr} has the following form
\begin{equation}\label{Lambda}
\Lambda=\alpha_1\,J_1|_{\Theta=0}+\alpha_2*J_1+\alpha_2^2\,J_2+\alpha_2(1+\alpha_1)*J_2-\alpha_2(\beta_1 J_{susy}-\beta_2*J_{susy})\,,
\end{equation}
where the coefficients $\alpha_{1,2}$ and $\beta_{1,2}$ are the same as in \eqref{Lcoset}.

Now note that in the limit \eqref{eps} the Lax connection reduces to the Hodge dual of the conserved current \eqref{Jcurrent}
\begin{equation}\label{e0}
\lim_{\epsilon\rightarrow 0}\, \frac{1}{\epsilon}\Lambda=*J\,.
\end{equation}
Comparison of eq. \eqref{e0} with \eqref{J} suggests a non--straightforward relation between the two connections via the following gauge transformation depending on the spectral parameter and accompanied by the shift in the $X$--dependence of $\Lambda$ \footnote{The shift $X\rightarrow X+i\upsilon\Gamma^A\vartheta$ in $\Lambda$ is required since $\Lambda$ and $L$ are constructed in different coordinate systems (see Section 2 of \cite{Cagnazzo:2009zh} for more details about the choice of the coordinate basis).}
\begin{equation}\label{LLambda}
\Lambda(X^M+i\upsilon\Gamma^M\vartheta,\Theta)=\mathcal G_{\mathbf x}(X,\Theta)\,L(X,\Theta)\,\mathcal G_{\mathbf x}^{-1}(X,\Theta)-d{\mathcal {G_{\mathbf x}G_{\mathbf x}}}^{-1}(X,\Theta)\,,
\end{equation}
where both sides are truncated to quadratic order in fermions, $\Gamma^M=\Gamma^A\,e_A{}^M(X)$ and $\mathcal G_{\mathbf x}(X,\Theta)$ is an isometry supergroup element depending on the spectral parameter $\mathbf x$, which in the exponential parametrization has the following form
\begin{equation}\label{gvtheta}
\mathcal G_{\mathbf x}(X^A,\Theta)=\,e^{(X^A+i\upsilon\Gamma^A(1-V^2)\vartheta) P_A}\,e^{Q V\vartheta\,}h(i_\Delta\,\Omega_0^{AB})\,,
\end{equation}
where
$$
h(i_\Delta\,\Omega_0^{AB})=e^{-\frac{1}{2}i\upsilon\Gamma^C(1-V^2)\vartheta\,\Omega_{0C}{}^{AB}(X,\vartheta) M_{AB}}
$$
 is a compensating gauge transformation in the stability subgroup $H$ (\emph{i.e.} $SO(1,3)\times U(3)$ or $SO(1,1)\times SO(2)$) of the superisometry group $G$.
\if{}
\\
\\
\textbf{SKETCH OF THE PROOF OF \eqref{LLambda}. }First we should check that \eqref{LLambda} holds for terms with $\nabla\Theta$. To this end it is enough to consider the gauge transformation of
$$
L=(1+\alpha_1)e^A(X)P_A+\alpha_2* e^A(X)P_A+(2i\alpha_2\upsilon\Gamma^A*\nabla\vartheta+2i\alpha_2\upsilon\Gamma^A\Gamma_{11}\nabla\vartheta)P_A+\cdots\,.
$$
Applying the $e^{(X+\Delta)P}$ part of \eqref{gvtheta} to the above terms we get
\begin{equation}\label{1}
\hat L=(1+\alpha_1)e^A(X)K_A(X+\Delta)-e^A(X+\Delta)K_A(X+\Delta)+\alpha_2* e^A(X)K_A(X+\Delta)+(2i\alpha_2\upsilon\Gamma^A*\nabla\vartheta+2i\alpha_2\upsilon\Gamma^A\Gamma_{11}\nabla\vartheta)K_A+\cdots\,
\end{equation}
where $\Delta=i\upsilon\Gamma^A(1-V^2)\vartheta e_A^M(X)$. At this stage of the proof we do not need to consider the variation of $K_A(X+\Delta)$ and can just replace it with $K_A(X)$. Then eq. \eqref{1} takes the form
\begin{eqnarray}\label{2}
&\hat L=\alpha_1e^A(X)K_A-\nabla(\upsilon\Gamma^A(-\alpha_1+\alpha_2\Gamma_{11})\vartheta)K_A+\alpha_2* e^A(X)K_A+(2i\alpha_2\upsilon\Gamma^A*\nabla\vartheta+2i\alpha_2\upsilon\Gamma^A\Gamma_{11}\nabla\vartheta)K_A+\cdots\,&\nonumber\\
&=\alpha_1(e^A(X)+i\nabla(\upsilon\Gamma^A\vartheta))K_A+\alpha_2(* e^A(X)+2i\upsilon\Gamma^A*\nabla\vartheta)K_A+i\alpha_2(\Theta\Gamma^A\Gamma_{11}\nabla\Theta)K_A+\cdots\,&\\
&=\alpha_1e^A(X^M+i\upsilon\Gamma^M\vartheta)K_A+\alpha_2(* e^A(X^M+i\upsilon\Gamma^M\vartheta)+i\Theta\Gamma^A*\nabla\Theta +i\Theta\Gamma^A\Gamma_{11}\nabla\Theta)K_A+\cdots\,&\nonumber\\
&=\Lambda(X^M+i\upsilon\Gamma^M\vartheta,\Theta)&\nonumber
\end{eqnarray}
\fi

Actually, as an independent derivation procedure, alternative to that described in Section \ref{comleteL}, we also got the form of the terms in $L'$ quadratic in fermions in the Lax connection \eqref{L'} by performing the inverse gauge transformation, from $\Lambda$ to $L$.

Note that in contrast to \eqref{Lcoset} and \eqref{L'} the Lax connection \eqref{Lambda} is not directly invariant under the $\mathbb Z_4$--transformations \eqref{Z4auto}, \eqref{Ox} and \eqref{Oalpha}. In particular, its first ($\alpha_1$--dependent) term acquires the shift $-2e^{A}\Omega(gP_Ag^{-1}|_{\vartheta=0})$. To get back $\Lambda$ in its initial form the $\mathbb Z_4$--transformed Lax connection
$$
\Omega(\Lambda({\tt x}))=\Omega^{-1}\Lambda(\frac{1}{\tt x})\, \Omega
$$
should undergo a compensating gauge transformation $\mathcal G_\Omega$ and one finds
$$
\Lambda=\mathcal G_\Omega\,\Omega(\Lambda)\,\mathcal G^{-1}_\Omega-\mathcal G_\Omega d\mathcal G^{-1}_\Omega, \qquad {\rm where} \qquad \mathcal G_\Omega=\mathcal G_{\mathbf x}\Omega^{-1} \mathcal G_{\mathbf x}^{-1}\Omega\,,
$$
$\mathcal G_{\mathbf x}(X,\Theta)$ is the same as in \eqref{LLambda} and $\Lambda$ is evaluated at $X^M+i\upsilon\Gamma^M\vartheta$. Of course, this gauge transformation, which also affects the spectral parameter $\tt x$, is nothing but a different form of the relation \eqref{LLambda} taking into account the $\mathbb Z_4$--invariance of $L$.

\subsection{Lax connection and kappa--symmetry}
The Green--Schwarz formulation of the superstring is invariant under the local fermionic transformations of the target--space coordinates $Z^{\mathcal M}=(X^M,\Theta^\mu)$ which satisfy the following properties
\begin{equation}\label{kappastring}
\delta_\kappa Z^{\mathcal M}\,{\mathcal E}_{\mathcal M}{}^{  \alpha}=
\frac{1}{2}(1+\Gamma)^{  \alpha}_{~ \beta}\,
\kappa^{ \beta}(\xi),\qquad {  \alpha}=1,\cdots, 32
\end{equation}
\begin{equation}\label{kA}
\hskip-2.5cm\delta_\kappa Z^{\mathcal M}\,{\mathcal E}_{\mathcal M}{}^A=0,
\qquad   A=0,1,\cdots,9
\end{equation}
where $\kappa^{ \alpha}(\xi)$ is a 32--component spinor
parameter, $\frac{1}{2}(1+\Gamma)^{
\alpha}_{~ \beta}$ is a spinor projection matrix with
\begin{equation}\label{gbs}
\Gamma=\frac{1}{2\,\sqrt{-\det{g_{ij}}}}\,\epsilon^{ij}\,{\mathcal
E}_{i}{}^A\,{\mathcal E}_{j}{}^B\,\Gamma_{AB}\,\Gamma_{11}, \qquad
\Gamma^2=1\,,
\end{equation}
and $g_{ij}$ is an induced worldsheet metric.

The string equations of motion \eqref{Psi} and \eqref{B} transform into each other under the kappa--symmetry variations. Since the condition for the Lax connection to have zero--curvature is in one to one correspondence with the equations of motion, it is natural to assume that on the mass--shell the Lax connection should be invariant under the kappa--symmetry transformations, at least, modulo a gauge transformation. This is indeed so in the case of the supercoset sigma--models (see e.g. \cite{Arutyunov:2008if}). The explicit check that also the non--coset Lax connection \eqref{eq:Lax-connection}, \eqref{L'} possesses this property would be somewhat cumbersome, but fortunately one should not do this, because there is a simple generic proof that makes this fact evident. Indeed, since any Lax curvature depends on the left--hand--sides of the equations of motion (as \emph{e.g.} in \eqref{R01} and \eqref{eq:dLAdS2}), its variation under \eqref{kappastring} and \eqref{kA} also depends on the field equations and hence vanishes on--shell. This means that kappa--variation of the Lax connection leaves its curvature zero and, therefore, the kappa--transformed Lax connection is related to the initial one by a corresponding infinitesimal gauge transformation taking values in the isometry superalgebra.

\section{Conclusion}
We have constructed the zero--curvature Lax connections for Green--Schwarz superstrings in $AdS_4\times CP^3$ and $AdS_2\times S^2\times T^6$ superbackgrounds which generalize the corresponding supercoset sigma--model Lax connections with contributions due to the physical world--sheet fermionic modes associated with non--supersymmetric directions of the target superspaces. We have shown that the contribution of the non--coset fermions does not spoil the important property of the Lax connections being $\mathbb Z_4$--invariant and demonstrated how the obtained Lax connections are related via gauge transformations to the Lax connections constructed in \cite{Sorokin:2010wn,Sorokin:2011rr} with the use of an alternative (Noether--current) prescription.

Having at hand Lax connections which include the contribution of non--coset worldsheet modes one can address the problem of how these modify the algebraic curve and Bethe ansatz equations for the full superstring theory in these backgrounds. This should lead to a more general approach to integrability of Green--Schwarz superstrings which does not rely on having a supercoset sigma--model description of the string.

The terms in the Lax connections containing the non--supercoset fermions $\upsilon$ have been computed to the second order in $\upsilon$. An interesting and important open problem is to understand the structure of the Lax connections to all the orders in the non--coset fermions. Presumably, the series in $\upsilon$ (eq. \eqref{L'}) would converge into covariant expressions in terms of background superfields (supervielbeins, connection etc.) as happens for the supercoset Lax connections expressed in terms of the superisometry Cartan forms.

In the process of the construction of the Lax connections we have obtained the form of the superfield quantities (supervielbeins, connection, NS--NS field strength and dilatino) that describe the $AdS_2\times S^2\times T^6$ superbackground to all orders in the supercoset fermions and to the second order in the non--coset ones. We have also obtained the explicit form of the spin connection of the $AdS_4\times CP^3$ superspace (to the second order in $\upsilon$) which was left out in \cite{Gomis:2008jt}. In contrast to the Green--Schwarz formulation, the knowledge of the form of the spin connection is required, for instance, for the pure spinor description of superstrings in curved superbackgrounds \cite{Berkovits:2001ue} and, in particular, is needed for extending to the full $AdS_4\times CP^3$ superspace the supercoset pure--spinor sigma--model of \cite{Fre:2008qc,Bonelli:2008us,D'Auria:2008cw}. With some more efforts, which will be made elsewhere, one can also compute the form of the superfield strengths $F_2$ and $F_4$ of the RR fluxes in type IIA $AdS_2\times S^2\times T^6$ superbackgrounds and corresponding quantities describing this superbackground in the type IIB case. These are also required for the construction of the pure spinor string action and for studying D--branes in these superbackgrounds.

Finally a detailed knowledge of string theory in $AdS_2\times S^2\times T^6$ might shed light on the corresponding $AdS_2/CFT_1$  holographic duality which so far has not been well understood. This correspondence is especially important due to the relation to black holes in $D=4$ that have an $AdS_2\times S^2$ near--horizon geometry. The integrable string model considered here could for example be used to make predictions for anomalous dimensions of operators on the gauge--theory side which could be compared to those computed in a given candidate dual theory.

\subsection*{Acknowledgements}
The authors are grateful to Igor Bandos, Arkady Tseytlin and Konstantin Zarembo for valuable discussions and comments. Work of A.C. and D.S. was partially supported by the INFN Special Initiative TV12. D.S. was also supported in part by the MIUR-PRIN contract 2009-KHZKRX and the grant FIS2008-1980 of the Spanish MICINN. D.S. is grateful to the Department of Theoretical Physics of the Basque Country University for hospitality and the IKERBASQUE Foundation for a visiting fellowship. The research of L.W. was supported in part by NSF grants PHY-0555575 and PHY-0906222.

\newpage

\appendix

\section{Supercoset description of $AdS_4\times CP^3$ and $AdS_2\times S^2\times T^6$}
\label{sec:supercosets}

\subsection{$OSp(6|4)$ and $PSU(1,1|2)\rtimes E(6)$ superalgebras}
The superisometry group of $AdS_4\times CP^3$ is $OSp(6|4)$ while the (local) superisometry group of $AdS_2\times S^2\times T^6$ is $PSU(1,1|2)\rtimes E(6)$ (the semi-direct product\footnote{It is the semi-direct rather than the direct product since the fermionic generators of $PSU(1,1|2)$ transform under a $U(1)\subset SO(6)\subset E(6)$ as follows from (\ref{eq:PQcom}), see also \cite{Sorokin:2011rr}.} of $PSU(1,1|2)$ with generators $(P_{\underline a},M_{\underline{ab}},Q)$ and the Euclidean group in six dimensions with generators $(P_{a'},M_{a'b'})$). The commutation relations of both the $OSp(6|4)$ and $PSU(1,1|2)\rtimes E(6)$ superalgebras can be written in 10D notation as
\begin{eqnarray}
{}[P_A,P_B]=-\frac{1}{2}R_{AB}{}^{CD}M_{CD}\,,\qquad[M_{AB},P_C]=\eta_{AC}P_B-\eta_{BC}P_A\nonumber\\
{}[M_{AB},M_{CD}]=\eta_{AC}M_{BD}+\eta_{BD}M_{AC}-\eta_{BC}M_{AD}-\eta_{AD}M_{BC}\,,
\end{eqnarray}
where $R_{AB}{}^{CD}$ is the Riemann tensor of $AdS_4\times CP^3$ or $AdS_2\times S^2\times T^6$ respectively, which are given in Section \ref{sec:notation}, while the commutation relation involving fermionic generators are
\begin{eqnarray}
[P_A,Q]=\frac{i}{R}Q\gamma_\star \Gamma_A\mathcal P\,\qquad[M_{AB},Q]=-\frac{1}{2}Q\Gamma_{AB}\mathcal P\,,\label{eq:PQcom}\\
\{Q,Q\}=2i(\mathcal P\Gamma^A\mathcal P)\,P_A+\frac{R}{4}(\mathcal P\Gamma^{AB}\gamma_\star \mathcal P)R_{AB}{}^{CD}M_{CD}\,.
\end{eqnarray}
In these expressions $\gamma_\star $, $\mathcal P$ and $R$ are given by
\begin{equation}\label{gammasA}
\begin{array}{c|cc}
& OSp(6|4) & PSU(1,1|2)\rtimes E(6)\\
\hline&&\\
\gamma_\star  & \gamma^5 & \Gamma^{01}\gamma^7\\
\mathcal P & \mathcal P_{24} & \mathcal P_8\\
R & 2R_{AdS_4} & 2R_{AdS_2}
\end{array}\,,
\end{equation}
where
$$
\gamma^5=i\Gamma^{0123}\, \qquad {\rm and}\qquad\gamma^7=i\Gamma^{456789}\,.
$$

\subsection{Supercoset geometry}\label{SG}
The supercoset corresponding to $AdS_4\times CP^3$ with 24 fermionic directions is
\begin{equation}
\frac{OSp(6|4)}{SO(1,3)\times U(3)}\,,
\end{equation}
while the supercoset corresponding (locally) to $AdS_2\times S^2\times T^6$ with 8 fermionic directions is
\begin{equation}
\frac{PSU(1,1|2)\rtimes E(6)}{SO(1,1)\times SO(2)\times SO(6)}\,,
\end{equation}
where $E(6)$ is the Euclidean group in six dimensions. The supercosets are parametrized by bosonic coordinates $X^M$ and fermionic coordinates $\vartheta=\mathcal P\Theta$.  The corresponding supercoset geometries are described by Cartan forms satisfying the Maurer--Cartan equation
\begin{equation}
K(X,\vartheta)=g^{-1}dg(X,\vartheta)=\frac{1}{2}\Omega_0^{AB}M_{AB}+E^AP_A+QE\,,\qquad dK=KK\,.
\end{equation}
For instance, if one chooses the isometry supergroup element in the form $g(X,\vartheta)=e^{P_AX^A}e^{Q\vartheta}$, the supervielbeins and spin connection are given by
\begin{eqnarray}
E^A&=&e^A+2i\vartheta\Gamma^A\frac{\cosh{\mathcal M}-1}{{\mathcal M}^2}D\vartheta\,,\nonumber\\
E^\alpha&=&\left(\frac{\sinh{\mathcal M}}{\mathcal M}D\vartheta\right)^\alpha\,,\nonumber\\
\Omega_0^{AB}&=&\omega^{AB}+\frac{R}{2}\vartheta\Gamma^{CD}\gamma_\star \frac{\cosh{\mathcal M}-1}{{\mathcal M}^2}D\vartheta\,R_{CD}{}^{AB}\,,
\label{eq:coset-geom}
\end{eqnarray}
where
\begin{equation}
\mathcal M^2=
-\frac{2}{R}(\mathcal P\gamma_\star \Gamma_A\vartheta)(\vartheta\Gamma^A\mathcal P)
-\frac{R}{8}R_{AB}{}^{CD}(\Gamma_{CD}\vartheta)(\vartheta\Gamma^{AB}\gamma_\star )\,,
\end{equation}
while
\begin{equation}
D\vartheta=\mathcal P(d-\frac{1}{4}\omega^{AB}\Gamma_{AB}+\frac{i}{R}e^A\,\gamma_\star \Gamma_A)\vartheta
\end{equation}
is the Killing--spinor derivative.

From the Maurer-Cartan equation one also finds the supercoset torsion
\begin{eqnarray}
\nabla E^A\equiv dE^A+E^B\Omega_{0B}{}^{A}&=&-iE\Gamma^AE,\\
\nabla E\equiv(d-\frac{1}{4}\Omega_0^{AB}\Gamma_{AB})E&=&\frac{i}{R}E^A\,(\mathcal P\gamma_\star \Gamma_AE)
\end{eqnarray}
and curvature
\begin{equation}
d\Omega_0^{AB}+\Omega_0^{AC}\Omega_{0C}{}^B=(\frac{1}{2}E^DE^C-\frac{R}{4}E\Gamma^{CD}\gamma_\star E)R_{CD}{}^{AB}\,.
\end{equation}

\section{Superstring equations of motion on $AdS_4\times CP^3$ to the second order in $\upsilon$}\label{eom}
To derive the form of the corrections \eqref{L'} to the supercoset Lax connection due to the non--coset fermionic modes $\upsilon$ we have had to compute the corresponding corrections to the superstring field equations \eqref{PsiB}, which have the following form
\begin{eqnarray}\label{Psi2}
\Psi_0&\equiv& i*E^A\,(\Gamma_AE)-iE^A\,(\Gamma_A\Gamma_{11}E)
\nonumber\\
&=&
-i*(2i\upsilon\Gamma^AE+i\upsilon\Gamma^AD\upsilon)\,(\Gamma_AE)
+\frac{i}{R}*E^A\,\upsilon\gamma^5\upsilon(\Gamma_AE)
+\frac{i}{2R}*E^A\,\upsilon\Gamma^b\gamma^5\upsilon\,(\Gamma_A\Gamma_bE)
\nonumber\\
&&{}
-\frac{i}{R}*E^A\,\upsilon\Gamma^b\gamma^7\upsilon\,(\Gamma_A\Gamma_b\Gamma_{11}E)
-\frac{i}{2R}*E^A\,\upsilon\upsilon\,(\Gamma_A\gamma_5E)
-\frac{i}{R}E^A\,\upsilon\gamma^5\upsilon\,(\Gamma_A\Gamma_{11}E)
\nonumber\\
&&{}
+i(2i\upsilon\Gamma^AE+i\upsilon\Gamma^AD\upsilon)\,(\Gamma_A\Gamma_{11}E)
-\frac{i}{2R}E^A\,\upsilon\Gamma^b\gamma^5\upsilon\,(\Gamma_A\Gamma_{11}\Gamma_bE)
\nonumber\\
&&{}
-\frac{i}{R}E^A\,\upsilon\Gamma^b\gamma^7\upsilon\,(\Gamma_A\Gamma_bE)
+\frac{i}{2R}E^A\,\upsilon\upsilon\,(\Gamma_A\gamma_7E)
-i*\hat E^A\,(\Gamma_AD\upsilon)
\nonumber\\
&&{}
+i\hat E^A\,(\Gamma_A\Gamma_{11}D\upsilon)
+\frac{1}{R}*\hat E^A\,\hat E_A\,(\gamma^5\upsilon)
+\frac{1}{R}\hat E^A\hat E^B\,(\Gamma_{AB}\gamma_7\upsilon)
\end{eqnarray}
and
\begin{eqnarray}\label{B2}
\mathcal B_0^A&\equiv&\nabla*E^A-iE\Gamma^A\Gamma_{11}E
\nonumber\\
&=&
-\nabla*\left[2i\upsilon\Gamma^AE+i\upsilon\Gamma^AD\upsilon-\frac{1}{R}\upsilon\gamma^5\upsilon E^A\right]
+\frac{2}{R}*E^A D\upsilon\, \gamma^5 \upsilon
+2iE\Gamma^A\Gamma_{11}D\upsilon
\nonumber\\
&&{}
+iD\upsilon\Gamma^A\Gamma_{11} D\upsilon
+2iE\Gamma^A\Gamma_{11} \left(-\frac{1}{2R}\upsilon\Gamma^b\gamma^5\upsilon\,\Gamma_b+\frac{1}{R}\upsilon\Gamma^b\gamma^7\upsilon\,\Gamma_b\Gamma_{11}+\frac{1}{2R}\upsilon\upsilon\,\gamma_5\right)E
\nonumber\\
&&{}
-\frac{2}{R}*\hat E^{b'}\delta^A_{a'}\upsilon\Gamma^{a'}{}_{b'}\gamma_5E
+\frac{2}{R}*E^b\delta^A_a\,\upsilon\Gamma^a{}_b\gamma_5D\upsilon
+\frac{4i}{R^2}*E^{b'}E^c\delta^A_{a'}\,\upsilon\Gamma^{a'}{}_{b'c}\upsilon
\nonumber\\
&&{}
-\frac{4}{R}\hat E^B\,E\Gamma^A{}_B \gamma^7\upsilon
-\frac{4}{R}E^B\,D\upsilon\,\Gamma^A{}_B\gamma^7\upsilon
+\frac{6i}{R^2}E^cE^b\upsilon\Gamma_{abc}\Gamma_{11}\upsilon\,,
\end{eqnarray}
where $\hat E^A=E^A+2i\upsilon\Gamma^AE$.

\section{Curvature of the Lax connection}
\label{sec:dL}
The curvature of the Lax connection (\ref{eq:Lax-connection}), \eqref{L'} computed to quadratic order in the non--coset fermions $\upsilon$ can be split into three pieces corresponding to the generators in the superalgebra, $M_{AB}$, $P_A$ and $Q$:
\begin{equation}
dL-LL=(dL-LL)_M+(dL-LL)_P+(dL-LL)_Q\,.
\end{equation}
With a bit of work one finds that, to order $\upsilon^2$,
\begin{eqnarray}
\lefteqn{(dL-LL)_M=}\nonumber\\
&&\frac{\alpha_2}{4}\Big[
i*\hat E^A\,\upsilon\Gamma^{CD}\Gamma_AV^2E
+i*E^A\,\upsilon\Gamma^{CD}\Gamma_AV^2\nabla\upsilon
-i\hat E^A\,\upsilon\Gamma^{CD}\Gamma_A\Gamma_{11}V^2E
\nonumber\\
&&{}
-iE^A\,\upsilon\Gamma^{CD}\Gamma_A\Gamma_{11}V^2\nabla\upsilon
+4\upsilon\Gamma^CV^2E\,\upsilon\Gamma^D\Gamma_{11}E
+4\upsilon\Gamma^CE\,\upsilon\Gamma^D\Gamma_{11}V^2E
\nonumber\\
&&{}
-\upsilon\Gamma^A\Gamma_{11}E\,E\Gamma^{CD}\mathcal P\Gamma_AV^2\upsilon
-\upsilon\Gamma^AE\,E\Gamma^{CD}\mathcal P\Gamma_A\Gamma_{11}V^2\upsilon
+\frac{i}{2}\nabla*E^A\,\upsilon\Gamma_A{}^{CD}V^2\upsilon
\nonumber\\
&&{}
-\frac{1}{2}E\Gamma^AE\,\upsilon\Gamma_A{}^{CD}\Gamma_{11}V^2\upsilon
+\frac{4}{R}*E^CE^A\,\upsilon\Gamma^DV^2\mathcal P\gamma_\star \Gamma_A\upsilon
-\frac{4}{R}E^CE^A\,\upsilon\Gamma^DV^2\Gamma_{11}\mathcal P\gamma_\star \Gamma_A\upsilon
\nonumber\\
&&{}
-\frac{\alpha_2}{R}*E^BE^A\,\upsilon\Gamma_B\mathcal P\gamma_\star \Gamma^{CD}\mathcal P\Gamma_A\Gamma_{11}\upsilon
-\frac{\alpha_2}{R}E^BE^A\,\upsilon\Gamma_B\mathcal P\gamma_\star \Gamma^{CD}\mathcal P\Gamma_A\upsilon
\Big]R_{CD}{}^{EF}M_{EF}\,,
\end{eqnarray}
where we have again introduced $\hat E^A=E^A+2i\upsilon\Gamma^AE$ to shorten the expressions. The terms in the Lax curvature proportional to $P_A$ are
\begin{eqnarray}
\lefteqn{(dL-LL)_P=}\nonumber\\
&&\alpha_2\Big[
\nabla*(\hat E^A+i\upsilon\Gamma^A\nabla\upsilon)
-iE\Gamma^A\Gamma_{11}E
-2iE\Gamma^A\Gamma_{11}\nabla\upsilon
-i\nabla\upsilon\Gamma^A\Gamma_{11}\nabla\upsilon
\nonumber\\
&&{}
-\frac{2}{R}E^B\,\upsilon\Gamma^A\Gamma_{11}\mathcal P\gamma_\star \Gamma_BE
-\frac{2}{R}E^B\,\upsilon\Gamma^A\Gamma_{11}\mathcal P\gamma_\star \Gamma_B\nabla\upsilon
-\frac{2}{R}*\hat E^B\,E\Gamma^A\mathcal P\gamma_\star \Gamma_B\upsilon
\nonumber\\
&&{}
-\frac{2}{R}*E^B\,\nabla\upsilon\Gamma^A\mathcal P\gamma_\star \Gamma_B\upsilon
+\frac{2}{R}\hat E^B\,E\Gamma^A\mathcal P\gamma_\star \Gamma_B\Gamma_{11}\upsilon
+\frac{2}{R}E^B\,\nabla\upsilon\Gamma^A\mathcal P\gamma_\star \Gamma_B\Gamma_{11}\upsilon
\nonumber\\
&&{}
-\frac{2}{R}*E^B\,\upsilon\Gamma^A\mathcal P\gamma_\star \Gamma_B\nabla\upsilon
+\frac{2i}{R}E\Gamma^BE\,\upsilon\Gamma^A\Gamma_{11}\mathcal P\gamma_\star \Gamma_B\upsilon
-\frac{2i}{R}E\Gamma^B\Gamma_{11}E\,\upsilon\Gamma^A\mathcal P\gamma_\star \Gamma_B\upsilon
\nonumber
\end{eqnarray}
\begin{eqnarray}
&&{}
+\frac{iR}{16}E\Gamma_{DE}\gamma_\star E\,\upsilon\Gamma^A\Gamma_{11}\Gamma^{BC}\upsilon\,R_{BC}{}^{DE}
-\frac{i}{8}E^FE^D\,\upsilon\Gamma^{ABC}\Gamma_{11}\upsilon\,R_{BCDF}
\nonumber\\
&&{}
-\frac{i}{4}E^FE^D\,\upsilon\Gamma_D{}^{BC}\Gamma_{11}\upsilon\,R_{BCF}{}^A
+\frac{i}{4}E^B*E^F\,\upsilon\Gamma_F{}^{DE}\upsilon\,R_{DEB}{}^A
-\frac{2i}{R}\upsilon\Gamma^B\Gamma_{11}E\,E\Gamma^A\mathcal P\gamma_\star \Gamma_B\upsilon
\nonumber\\
&&{}
-\frac{2i}{R}\upsilon\Gamma^BE\,E\Gamma^A\mathcal P\gamma_\star \Gamma_B\Gamma_{11}\upsilon
-\frac{2}{R}(\nabla*E^B-iE\Gamma^B\Gamma_{11}E)\,\upsilon\Gamma^A\mathcal P\gamma_\star \Gamma_B\upsilon
\Big]\,P_A\,.
\end{eqnarray}
Finally the terms proportional to $Q$ in the Lax curvature become
\begin{eqnarray}
\lefteqn{(dL-LL)_Q=}
\nonumber\\
&&\alpha_2\frac{i}{R}\Big[
(\hat E^A+i\upsilon\Gamma^A\nabla\upsilon)\,QV^\dagger\gamma_\star \Gamma_A\Gamma_{11}E
+\hat E^A\,QV^\dagger\gamma_\star \Gamma_A\Gamma_{11}\nabla\upsilon
\nonumber\\
&&{}
-*(\hat E^A+i\upsilon\Gamma^A\nabla\upsilon)\,QV^\dagger\gamma_\star \Gamma_AE
-*\hat E^A\,QV^\dagger\gamma_\star \Gamma_A\nabla\upsilon
\nonumber\\
&&{}
-\frac{i}{R}\hat E^A\hat E^B\,QV^\dagger\gamma_\star \Gamma_A\mathcal P\gamma_\star \Gamma_B\Gamma_{11}\upsilon
-\frac{i}{R}*\hat E^A\hat E^B\,QV^\dagger\gamma_\star \Gamma_A\mathcal P\gamma_\star \Gamma_B\upsilon
\nonumber\\
&&{}
-\frac{2}{R}*E^B\,E\Gamma^A\mathcal P\gamma_\star \Gamma_B\upsilon\,QV^\dagger\gamma_\star \Gamma_A\upsilon
+\frac{2}{R}*E^B\,\upsilon\Gamma^A\mathcal P\gamma_\star \Gamma_B\upsilon\,QV^\dagger\gamma_\star \Gamma_AE
\nonumber\\
&&{}
-\frac{2}{R}*E^B\,\upsilon\Gamma^AE\,QV^\dagger\gamma_\star \Gamma_A\mathcal P\gamma_\star \Gamma_B\upsilon
+\frac{2}{R}E^B\,E\Gamma^A\mathcal P\gamma_\star \Gamma_B\Gamma_{11}\upsilon\,QV^\dagger\gamma_\star \Gamma_A\upsilon
\nonumber\\
&&{}
+\frac{2}{R}E^B\,\upsilon\Gamma^A\Gamma_{11}\mathcal P\gamma_\star \Gamma_B\upsilon\,QV^\dagger\gamma_\star \Gamma_AE
+\frac{2}{R}E^B\,\upsilon\Gamma^AE\,QV^\dagger\gamma_\star \Gamma_A\mathcal P\gamma_\star \Gamma_B\Gamma_{11}\upsilon
\nonumber\\
&&{}
-\frac{1}{R}E^B\,\upsilon\Gamma^A\Gamma_{11}\mathcal P\gamma_\star \Gamma_BE\,QV^\dagger\gamma_\star \Gamma_A\upsilon
-\frac{1}{R}E^B\,\upsilon\Gamma^A\mathcal P\gamma_\star \Gamma_BE\,QV^\dagger\gamma_\star \Gamma_A\Gamma_{11}\upsilon
\nonumber\\
&&{}
-\frac{1}{R}E^B\,\upsilon\Gamma_AE\,QV^\dagger\gamma_\star \Gamma_B\gamma_\star \mathcal P\Gamma^A\Gamma_{11}\upsilon
-\frac{1}{R}E^B\,\upsilon\Gamma^A\Gamma_{11}E\,QV^\dagger\gamma_\star \Gamma_B\gamma_\star \mathcal P\Gamma_A\upsilon
\nonumber\\
&&{}
+\frac{R}{16}(*E^C\,\upsilon\Gamma_C{}^{DE}\upsilon-E^C\,\upsilon\Gamma_C{}^{DE}\Gamma_{11}\upsilon)\,R_{DE}{}^{AB}\,QV^\dagger\Gamma_{AB}E
\Big]
\nonumber\\
&&{}
\nonumber\\
&&{}
+\alpha_2^2\frac{i}{R^2}\Big(
*E^B\,\upsilon\Gamma^AE\,(2QV\gamma_\star \Gamma_A\mathcal P\gamma_\star \Gamma_B\Gamma_{11}\upsilon-QV\gamma_\star \Gamma_B\gamma_\star \mathcal P\Gamma^A\Gamma_{11}\upsilon)
\nonumber\\
&&{}
+*E^B\,\upsilon\Gamma^A\Gamma_{11}E\,(2QV\gamma_\star \Gamma_A\mathcal P\gamma_\star \Gamma_B\upsilon-QV\gamma_\star \Gamma_B\gamma_\star \mathcal P\Gamma_A\upsilon)
\nonumber\\
&&{}
-E^B\,\upsilon\Gamma^AE\,(2QV\gamma_\star \Gamma_A\mathcal P\gamma_\star \Gamma_B\upsilon-QV\gamma_\star \Gamma_B\gamma_\star \mathcal P\Gamma^A\upsilon)
\nonumber\\
&&{}
-E^B\,\upsilon\Gamma^A\Gamma_{11}E\,(2QV\gamma_\star \Gamma_A\mathcal P\gamma_\star \Gamma_B\Gamma_{11}\upsilon-QV\gamma_\star \Gamma_B\gamma_\star \mathcal P\Gamma_A\Gamma_{11}\upsilon)
\nonumber\\
&&{}
-\frac{R^2}{16}(*E^C\,\upsilon\Gamma_C{}^{DE}\upsilon-E^C\,\upsilon\Gamma_C{}^{DE}\Gamma_{11}\upsilon)\,R_{DE}{}^{AB}\,QV\Gamma_{AB}\Gamma_{11}E
\nonumber\\
&&{}
-\frac{R^2}{16}(*E^C\,\upsilon\Gamma_C{}^{DE}\Gamma_{11}\upsilon-E^C\,\upsilon\Gamma_C{}^{DE}\upsilon)\,R_{DE}{}^{AB}\,QV\Gamma_{AB}E
\Big)
\nonumber\\
&&{}
\nonumber\\
&&{}
-\alpha_2\frac{i}{R}\Big(
\nabla*\hat E^A-iE\Gamma^A\Gamma_{11}E
-2iE\Gamma^A\Gamma_{11}\nabla\upsilon
-\frac{2}{R}E^B\,\upsilon\Gamma^A\Gamma_{11}\mathcal P\gamma_\star \Gamma_BE
\nonumber\\
&&{}
+\frac{2}{R}E^B\,E\Gamma^A\mathcal P\gamma_\star \Gamma_B\Gamma_{11}\upsilon
-\frac{2}{R}*E^B\,E\Gamma^A\mathcal P\gamma_\star \Gamma_B\upsilon
\Big)\,QV^\dagger\gamma_\star \Gamma_A\upsilon
\,,
\end{eqnarray}
where we've used the fact that
\begin{equation}
2(\Gamma^A\Gamma_{11}E)_\alpha(\Gamma_AE)_\beta
+2(\Gamma_AE)_\alpha(\Gamma^A\Gamma_{11}E)_\beta
+(\Gamma^A\Gamma_{11})_{\alpha\beta}\,E\Gamma_AE
+(\Gamma^A)_{\alpha\beta}\,E\Gamma_A\Gamma_{11}E=0
\end{equation}
and
\begin{eqnarray}
\upsilon\Gamma^A\nabla\upsilon\,QV^\dagger\gamma_\star \Gamma_A\Gamma_{11}E
+\upsilon\Gamma^A\Gamma_{11}\nabla\upsilon\,QV^\dagger\gamma_\star \Gamma_AE
+E\Gamma^A\Gamma_{11}\nabla\upsilon\,QV^\dagger\gamma_\star \Gamma_A\upsilon
\nonumber\\
{}+E\Gamma^A\nabla\upsilon\,QV^\dagger\gamma_\star \Gamma_A\Gamma_{11}\upsilon
+\upsilon\Gamma^A\Gamma_{11}E\,QV^\dagger\gamma_\star \Gamma_A\nabla\upsilon
+\upsilon\Gamma_AE\,QV^\dagger\gamma_\star \Gamma^A\Gamma_{11}\nabla\upsilon
=0
\end{eqnarray}
which follow from the basic Fierz identity (\ref{eq:Fierz1}).

\section{Gamma-matrix identities}
\label{sec:gammamatrices}

Some useful gamma-matrix identities are ($a=0,1,2,3$)
\begin{eqnarray}
\Gamma^{abc}&=&-i\varepsilon^{abcd}\Gamma_d\gamma^5\\
\Gamma^{ab}&=&-\frac{i}{2}\varepsilon^{abcd}\Gamma_{cd}\gamma^5\\
\Gamma^a&=&\frac{i}{6}\varepsilon^{abcd}\Gamma_{bcd}\gamma^5
\end{eqnarray}
and some useful identities involving the projection operators are
\begin{eqnarray}
\mathcal P_8\Gamma^{a'b'c'}\mathcal P_{24}&=&-3iJ^{[a'b'}\mathcal P_8\Gamma^{c']}\gamma_7\mathcal P_{24}\\
\mathcal P_8\Gamma^{a'b'}\mathcal P_8&=&iJ^{a'b'}\gamma^7\mathcal P_8\\
\Gamma^{a'}\mathcal P_8\Gamma_{a'}&=&2\mathcal P_{24}\\
\Gamma^{[a'}\mathcal P_8\Gamma^{b']}&=&\frac{1}{2}\mathcal P_{24}\Gamma^{a'b'}\mathcal P_{24}+\frac{i}{2}J^{a'b'}\gamma_7\mathcal P_{24}\\
\mathcal P_{24}(\delta_{a'}^{b'}+iJ_{a'}{}^{b'}\gamma_7)\Gamma_{b'}\mathcal P_{24}&=&0\\
\mathcal P_8(\delta_{a'}^{b'}-iJ_{a'}{}^{b'}\gamma_7)\Gamma_{b'}\mathcal P_{24}&=&0\,.
\end{eqnarray}

\subsection{Fierz identities}
The basic Fierz identity for the $D=11$ gamma-matrices we use says that
\begin{equation}
\Gamma^{\hat A}_{(\alpha\beta}(\Gamma_{\hat A\hat B})_{\gamma\delta)}=0\,,
\end{equation}
where $\hat A=0,\ldots,10$. In $D=10$ notation this becomes the identities
\begin{equation}
\Gamma^A_{(\alpha\beta}(\Gamma_A\Gamma_{11})_{\gamma\delta)}=0
\label{eq:Fierz1}
\end{equation}
and
\begin{equation}
\Gamma^A_{(\alpha\beta}(\Gamma_{AB})_{\gamma\delta)}+\Gamma^{11}_{(\alpha\beta}(\Gamma_{11}\Gamma_B)_{\gamma\delta)}=0\,.
\label{eq:Fierz2}
\end{equation}

We can also expand fermion bilinears in a Fierz basis as follows:
\begin{eqnarray}
\Theta^\alpha\,\Theta^\beta
&=&
\frac{1}{32}\mathcal C^{\alpha\beta}\,\Theta\Theta
-\frac{1}{32\cdot2}(\Gamma_{AB}\Gamma_{11})^{\alpha\beta}\,\Theta\Gamma^{AB}\Gamma_{11}\Theta
-\frac{1}{32\cdot3!}\Gamma_{ABC}^{\alpha\beta}\,\Theta\Gamma^{ABC}\Theta
\nonumber\\
&&{}
+\frac{1}{32\cdot3!}(\Gamma_{ABC}\Gamma_{11})^{\alpha\beta}\,\Theta\Gamma^{ABC}\Gamma_{11}\Theta
+\frac{1}{32\cdot4!}\Gamma_{ABCD}^{\alpha\beta}\,\Theta\Gamma^{ABCD}\Theta\,.
\end{eqnarray}
It will be useful to project this identity in various ways using our projection operators. In the $AdS_4\times CP^3$ case when $\upsilon=\mathcal P_8\upsilon$ we get
\begin{eqnarray}
\upsilon^\alpha\,\upsilon^\beta
&=&
\frac{1}{8}(\mathcal P_8\mathcal C)^{\alpha\beta}\,\upsilon\upsilon
+\frac{1}{8}(\gamma^5\mathcal P_8)^{\alpha\beta}\,\upsilon\gamma^5\upsilon
-\frac{1}{8}(\Gamma_a\gamma^5\mathcal P_8)^{\alpha\beta}\,\upsilon\Gamma^a\gamma^5\upsilon
\nonumber\\
&&{}
+\frac{1}{8}(\Gamma_a\gamma^7\mathcal P_8)^{\alpha\beta}\,\upsilon\Gamma^a\gamma^7\upsilon
-\frac{1}{16}(\Gamma_{ab}\gamma^7\mathcal P_8)^{\alpha\beta}\,\upsilon\Gamma^{ab}\gamma^7\upsilon
\nonumber\\
&&{}
-\frac{1}{32\cdot3!}(\mathcal P_8\Gamma_{a'b'c'}\mathcal P_8)^{\alpha\beta}\,\upsilon\Gamma^{a'b'c'}\upsilon
+\frac{1}{32\cdot3!}(\mathcal P_8\Gamma_{a'b'c'}\Gamma_{11}\mathcal P_8)^{\alpha\beta}\,\upsilon\Gamma^{a'b'c'}\Gamma_{11}\upsilon
\nonumber\\
&&{}
+\frac{1}{32\cdot3!}(\mathcal P_8\Gamma_a\Gamma_{b'c'd'}\mathcal P_8)^{\alpha\beta}\,\upsilon\Gamma^a\Gamma^{b'c'd'}\upsilon
\end{eqnarray}
and it follows from this expression that
\begin{eqnarray}
(\Gamma^{a'}\upsilon)^\alpha\,(\Gamma_{a'}\upsilon)^\beta
&=&
-\frac{1}{4}(\mathcal P_{24}\mathcal C)^{\alpha\beta}\,\upsilon\upsilon
-\frac{1}{4}(\gamma^5\mathcal P_{24})^{\alpha\beta}\,\upsilon\gamma^5\upsilon
-\frac{1}{4}(\Gamma_a\gamma^5\mathcal P_{24})^{\alpha\beta}\,\upsilon\Gamma^a\gamma^5\upsilon
\nonumber\\
&&{}
-\frac{1}{4}(\Gamma_a\gamma^7\mathcal P_{24})^{\alpha\beta}\,\upsilon\Gamma^a\gamma^7\upsilon
-\frac{1}{8}(\Gamma_{ab}\gamma^7\mathcal P_{24})^{\alpha\beta}\,\upsilon\Gamma^{ab}\gamma^7\upsilon\,.
\end{eqnarray}

Similarly we have in the $AdS_2\times S^2\times T^6$ case when $\upsilon=\mathcal P_{24}\upsilon$ that
\begin{eqnarray}
(\mathcal P_8\Gamma^{a'}\upsilon)^\alpha\,(\Gamma_{a'}\upsilon)^\beta
&=&
-\frac{1}{4}(\mathcal P_8\mathcal C)^{\alpha\beta}\,\upsilon\upsilon
-\frac{1}{4}(\mathcal P_8\gamma_5)^{\alpha\beta}\,\upsilon\gamma_5\upsilon
-\frac{1}{4}(\mathcal P_8\Gamma^{\underline a}\gamma_5)^{\alpha\beta}\,\upsilon\Gamma_{\underline a}\gamma_5\upsilon
\nonumber\\
&&{}
-\frac{1}{4}(\mathcal P_8\Gamma^{\underline a}\gamma_7)^{\alpha\beta}\,\upsilon\Gamma_{\underline a}\gamma_7\upsilon
-\frac{1}{8}(\mathcal P_8\Gamma_{\underline{ab}}\gamma_7)^{\alpha\beta}\,\upsilon\Gamma^{\underline{ab}}\gamma_7\upsilon
\nonumber\\
&&{}
-\frac{1}{4}(\mathcal P_8\Gamma_{\underline ab'}\gamma_5)^{\alpha\beta}\,\upsilon\Gamma^{\underline ab'}\gamma_5\upsilon
-\frac{1}{8}(\mathcal P_8\Gamma_{\underline{ab}c'})^{\alpha\beta}\,\upsilon\Gamma^{\underline{ab}c'}\upsilon\,.
\label{eq:AdS2Fierz}
\end{eqnarray}
These identities were used in many places in the calculation of the curvature of the Lax connection.

\section{Check of the closure of $H$ in $AdS_2\times S^2\times T^6$}
\label{sec:dH}
The NS--NS three-form superfield strength in the $AdS_2\times S^2\times T^6$ background is given by
\begin{eqnarray}
H&=&
-i\mathcal E^A\,\mathcal E\Gamma_A\Gamma_{11}\mathcal E
+i\mathcal E^B\mathcal E^A\,\mathcal E\Gamma_{AB}\Gamma_{11}\lambda
+\frac{1}{3!}\mathcal E^C\mathcal E^B\mathcal E^A\,H_{ABC}
\nonumber\\
&=&
-ic^2\mathcal E^A\,E\Gamma_A\Gamma_{11}E
-2ic^4\hat E^A\,E\Gamma_A\Gamma_{11}D\upsilon
-ic^4E^A\,D\upsilon\Gamma_A\Gamma_{11}D\upsilon
\nonumber\\
&&{}
+\frac{ic^4}{R}E^A\,\Big(
\upsilon\Gamma^B\gamma_\star \gamma_7\upsilon\,E\Gamma_{AB}\gamma_5E
-\upsilon\Gamma^B\gamma_\star \Gamma_{11}\upsilon\,E\Gamma_{AB}E
+\upsilon\Gamma^{AB}\gamma_\star \upsilon\,E\Gamma_B\Gamma_{11}E
\nonumber\\
&&{}
-\upsilon\Gamma^{\underline bc'}\Gamma_{11}\upsilon\,E\gamma_\star \Gamma_A\Gamma_{\underline bc'}E
+\frac{1}{2}\upsilon\Gamma^{\underline{bc}d'}\upsilon\,E\Gamma_{11}\gamma_\star \Gamma_A\Gamma_{\underline{bc}d'}E
\Big)
+\frac{2c^4}{R}\hat E^B\hat E^A\,E\Gamma_{AB}\Gamma_{11}\gamma_\star \upsilon
\nonumber\\
&&{}
+\frac{2c^4}{R}E^BE^A\,D\upsilon\Gamma_{AB}\Gamma_{11}\gamma_\star \upsilon
+\frac{c^6}{3!}E^CE^BE^A\,H_{ABC}
+\mathcal O(\upsilon^3)\,,
\end{eqnarray}
where $\mathcal E^A$ and $H_{ABC}$ are given is (\ref{eq:AdS2vielbeins}) and (\ref{eq:AdS2HABC}) respectively, and $\hat E^A=E^A+2i\upsilon\Gamma^AE$. We wish to demonstrate that this form is indeed closed.

After a bit of algebra using the torsion equation (\ref{TA}) and Fierz identities one finds
\begin{eqnarray}
\frac{1}{c^4}dH&=&
\frac{2i}{R}\hat E^B\,E\Gamma^AE\,E\Gamma_{AB}\Gamma_{11}\gamma_\star \upsilon
+\frac{2i}{R}\hat E^B\,E\Gamma_{AB}E\,E\Gamma^A\Gamma_{11}\gamma_\star \upsilon
-\frac{2i}{R}\hat E^B\,E\Gamma_{11}E\,E\Gamma_B\gamma_\star \upsilon
\nonumber\\
&&{}
+\frac{4i}{R}E^{b'}\,\upsilon\Gamma^{a'}E\,E\Gamma_{a'b'}\Gamma_{11}\gamma_\star \nabla\upsilon
+\frac{8i}{R}E^B\,E\Gamma^{a'}D\upsilon\,E\Gamma_{a'B}\Gamma_{11}\gamma_\star \upsilon
-\frac{2i}{R}E^{\underline a}\,E\Gamma_{\underline a}\Gamma_{11}E\,D\upsilon\gamma_\star \upsilon
\nonumber\\
&&{}
+\frac{2i}{R}E^{\underline b}\,\upsilon\Gamma^{\underline a}D\upsilon\,E\Gamma_{\underline a}\Gamma_{11}\gamma_\star \Gamma_{\underline b}E
-i\hat E^B\,(\Omega_{AB}-\Omega_{0AB})\,E\Gamma^A\Gamma_{11}E
-2iE^A\,D\upsilon\Gamma_A\Gamma_{11}\nabla D\upsilon
\nonumber\\
&&{}
-2iE^B\,(\Omega_{AB}-\Omega_{0AB})\,E\Gamma^A\Gamma_{11}D\upsilon
-\frac{4}{R}E^CE^B\,(\Omega_{AB}-\Omega_{0AB})\,E\Gamma^A{}_C\Gamma_{11}\gamma_\star \upsilon
\nonumber\\
&&{}
+\frac{4}{R^2}E^{a'}E^{\underline c}\,\upsilon\Gamma^{b'}E\,E\Gamma_{a'b'\underline c}\Gamma_{11}\upsilon
+\frac{i}{R}E^A\,\nabla\Big(
\upsilon\Gamma^B\gamma_\star \gamma_7\upsilon\,E\Gamma_{AB}\gamma_5E
-\upsilon\Gamma^B\gamma_\star \Gamma_{11}\upsilon\,E\Gamma_{AB}E
\nonumber\\
&&{}
+\upsilon\Gamma^{AB}\gamma_\star \upsilon\,E\Gamma_B\Gamma_{11}E
-\upsilon\Gamma^{\underline bc'}\Gamma_{11}\upsilon\,E\gamma_\star \Gamma_A\Gamma_{\underline bc'}E
+\frac{1}{2}\upsilon\Gamma^{\underline{bc}d'}\upsilon\,E\Gamma_{11}\gamma_\star \Gamma_A\Gamma_{\underline{bc}d'}E
\Big)
\nonumber\\
&&{}
+d\left(
\frac{2}{R}E^BE^A\,D\upsilon\Gamma_{AB}\Gamma_{11}\gamma_\star \upsilon
+\frac{c^2}{6}E^CE^BE^A\,H_{ABC}
\right)\,.
\end{eqnarray}
The terms in the first line vanish due to the Fierz identity (\ref{eq:Fierz2}). Using the expressions for $\Omega^{AB}$ and $H_{ABC}$ given in (\ref{eq:AdS2Omega}) and (\ref{eq:AdS2HABC}) and simplifying further this becomes
\begin{eqnarray}
&&{}
-\frac{1}{R^2}E^CE^{b'}\,E\Gamma^AE\,\upsilon\gamma_\star \Gamma_{b'}\Gamma_A\Gamma_{11}(1-P)\Gamma_C\gamma_\star \upsilon
+\ldots
\nonumber\\
&&{}
+\frac{2}{R^2}E^{b'}E^{\underline c}\,E\Gamma_{\underline c}\gamma_\star \Gamma_{a'}\upsilon\,E\Gamma_{11}\gamma_\star \Gamma_{b'}\Gamma_{a'}\upsilon
+\frac{2}{R^2}E^{b'}E^{\underline c}\,E\Gamma_{\underline c}\Gamma_{11}\gamma_\star \Gamma_{a'}\upsilon\,E\gamma_\star \Gamma_{b'}\Gamma_{a'}\upsilon
\nonumber\\
&&{}
-\frac{4}{R^2}E^{b'}E^{\underline c}\,E\Gamma^{a'}\upsilon\,E\Gamma_{b'\underline c}\Gamma_{11}\Gamma_{a'}\upsilon
-\frac{8}{R^2}E^BE^{\underline c}\,E\Gamma_{\underline c}\gamma_\star \gamma^{a'}\upsilon\,E\Gamma_B\Gamma_{11}\gamma_\star \Gamma_{a'}\upsilon
\nonumber\\
&&{}
-\frac{2}{R^2}E^{b'}E^{\underline c}\,E\Gamma_{\underline c}{}^{\hat a}E\,\upsilon\Gamma_{\hat ab'}\Gamma_{11}\upsilon
-\frac{2}{R^2}E^{b'}E^c\,E\gamma_7E\,\upsilon\Gamma_{cb'}\gamma_5\upsilon
-\frac{1}{R^2}E^{b'}E^{\underline c}\,E\Gamma^{\underline a}\Gamma_{11}E\,\upsilon\Gamma_{\underline{ca}b'}\upsilon
\nonumber\\
&&{}
-\frac{2}{R^2}E^BE^{\underline c}\,E\Gamma^{\underline a}\Gamma_{11}E\,\upsilon\gamma_\star \Gamma_{\underline{ca}B}\gamma_\star \upsilon
+\frac{2}{R^2}E^BE^{\underline c}\,E\Gamma^{\underline a}E\,\upsilon\gamma_\star \Gamma_{\underline{ca}B}\Gamma_{11}\gamma_\star \upsilon
\nonumber\\
&&{}
-\frac{1}{R^2}E^{b'}E^{\underline c}\,E\Gamma^{\underline a}E\,\upsilon\Gamma_{\underline{ca}b'}\Gamma_{11}\upsilon
+\frac{2i}{R}E^{a'}\,E\Gamma^{b\hat c}E\,\upsilon\Gamma_{a'b\hat c}\gamma_\star \Gamma_{11}\nabla\upsilon
+\frac{2i}{R}E^{\underline a}\,E\Gamma_{\underline{ab}}E\,\upsilon\gamma_\star \Gamma^{\underline b}\Gamma_{11}\nabla\upsilon
\nonumber\\
&&{}
+\frac{2i}{R}E^{\underline a}\,E\Gamma_{\underline{ab}}\gamma_5E\,\upsilon\gamma_\star \Gamma^{\underline b}\gamma_7\nabla\upsilon
+\frac{4i}{R}E^{b'}\,\upsilon\Gamma^{a'}E\,E\Gamma_{a'b'}\Gamma_{11}\gamma_\star \nabla\upsilon
+\frac{8i}{R}E^B\,E\Gamma^{a'}\nabla\upsilon\,E\Gamma_{a'B}\Gamma_{11}\gamma_\star \upsilon
\nonumber\\
&&{}
+\frac{4i}{R}E^{b'}\,\upsilon\gamma_\star \Gamma_{a'b'}E\,E\Gamma^{a'}\Gamma_{11}\nabla\upsilon
-\frac{3i}{R}E^{a'}\,E\Gamma^{\underline b}E\,\upsilon\gamma_\star \Gamma_{a'\underline b}\Gamma_{11}\nabla\upsilon
-\frac{3i}{R}E^{a'}\,E\Gamma^{\underline b}\Gamma_{11}E\,\upsilon\gamma_\star \Gamma_{a'\underline b}\nabla\upsilon
\nonumber\\
&&{}
-\frac{i}{R}E^{a'}\,E\Gamma^{\underline b}E\,\upsilon\Gamma_{a'\underline b}\gamma_\star \Gamma_{11}\nabla\upsilon
+\frac{i}{R}E^{a'}\,E\Gamma^{\underline b}\Gamma_{11}E\,\upsilon\Gamma_{a'\underline b}\gamma_\star \nabla\upsilon
-\frac{2i}{R}E^{\underline b}\,E\Gamma^{\underline a}E\,\upsilon\gamma_\star \Gamma_{\underline b}\Gamma_{\underline a}\Gamma_{11}\nabla\upsilon
\nonumber\\
&&{}
-\frac{2i}{R}E^{\underline b}\,E\Gamma^{\underline a}\Gamma_{11}E\,\upsilon\gamma_\star \Gamma_{\underline b}\Gamma_{\underline a}\nabla\upsilon
+\frac{2i}{R}E^A\,E\Gamma_{11}E\,\upsilon\gamma_\star \Gamma_A\nabla\upsilon
-\frac{2i}{R}E^A\,E\gamma_7E\,\upsilon\gamma_\star \gamma_5\Gamma_A\nabla\upsilon
\nonumber\\
&&{}
+\frac{4i}{R}E^{a'}\,E\gamma_7E\,\upsilon\gamma_\star \gamma_5\Gamma_{a'}\nabla\upsilon\,,
\end{eqnarray}
where the ellipsis in the first line denote three terms which, together with the previous term, cancel due to the Fierz identitity (\ref{eq:Fierz1}). Using the Fierz identity in (\ref{eq:AdS2Fierz})
the terms with two bosonic vielbeins can be seen to cancel and we are left with
\begin{eqnarray}
&&{}
\frac{2i}{R}E^{a'}\,E\Gamma^{b\hat c}E\,\upsilon\Gamma_{a'b\hat c}\gamma_\star \Gamma_{11}\nabla\upsilon
+\frac{2i}{R}E^{\underline a}\,E\Gamma_{\underline{ab}}E\,\upsilon\gamma_\star \Gamma^{\underline b}\Gamma_{11}\nabla\upsilon
+\frac{2i}{R}E^{\underline a}\,E\Gamma_{\underline{ab}}\gamma_5E\,\upsilon\gamma_\star \Gamma^{\underline b}\gamma_7\nabla\upsilon
\nonumber\\
&&{}
+\frac{4i}{R}E^{b'}\,\upsilon\Gamma^{a'}E\,E\Gamma_{a'b'}\Gamma_{11}\gamma_\star \nabla\upsilon
+\frac{8i}{R}E^B\,E\Gamma^{a'}\nabla\upsilon\,E\Gamma_{a'B}\Gamma_{11}\gamma_\star \upsilon
+\frac{4i}{R}E^{b'}\,\upsilon\gamma_\star \Gamma_{a'b'}E\,E\Gamma^{a'}\Gamma_{11}\nabla\upsilon
\nonumber\\
&&{}
-\frac{3i}{R}E^{a'}\,E\Gamma^{\underline b}E\,\upsilon\gamma_\star \Gamma_{a'\underline b}\Gamma_{11}\nabla\upsilon
-\frac{3i}{R}E^{a'}\,E\Gamma^{\underline b}\Gamma_{11}E\,\upsilon\gamma_\star \Gamma_{a'\underline b}\nabla\upsilon
-\frac{i}{R}E^{a'}\,E\Gamma^{\underline b}E\,\upsilon\Gamma_{a'\underline b}\gamma_\star \Gamma_{11}\nabla\upsilon
\nonumber\\
&&{}
+\frac{i}{R}E^{a'}\,E\Gamma^{\underline b}\Gamma_{11}E\,\upsilon\Gamma_{a'\underline b}\gamma_\star \nabla\upsilon
-\frac{2i}{R}E^{\underline b}\,E\Gamma^{\underline a}E\,\upsilon\gamma_\star \Gamma_{\underline b}\Gamma_{\underline a}\Gamma_{11}\nabla\upsilon
-\frac{2i}{R}E^{\underline b}\,E\Gamma^{\underline a}\Gamma_{11}E\,\upsilon\gamma_\star \Gamma_{\underline b}\Gamma_{\underline a}\nabla\upsilon
\nonumber\\
&&{}
+\frac{2i}{R}E^A\,E\Gamma_{11}E\,\upsilon\gamma_\star \Gamma_A\nabla\upsilon
-\frac{2i}{R}E^A\,E\gamma_7E\,\upsilon\gamma_\star \gamma_5\Gamma_A\nabla\upsilon
+\frac{4i}{R}E^{a'}\,E\gamma_7E\,\upsilon\gamma_\star \gamma_5\Gamma_{a'}\nabla\upsilon\,.
\end{eqnarray}
We now use the fact that
\begin{eqnarray}
\lefteqn{\frac{2i}{R}E^{\underline a}\,E\Gamma_{\underline{ab}}E\,\upsilon\gamma_\star \Gamma^{\underline b}\Gamma_{11}\nabla\upsilon
+\frac{2i}{R}E^{\underline a}\,E\Gamma_{\underline{ab}}\gamma_5E\,\upsilon\gamma_\star \Gamma^{\underline b}\gamma_7\nabla\upsilon}
\nonumber\\
&=&
\frac{2i}{R}E^{\underline a}\,E\Gamma_{\underline a}\Gamma_BE\,\upsilon\gamma_\star \Gamma^B\Gamma_{11}\nabla\upsilon
+\ldots
+\frac{2i}{R}E^{\underline a}\,E\gamma_5\Gamma_{\underline a}\Gamma_BE\,\upsilon\gamma_\star \Gamma^B\Gamma_{11}\gamma_5\nabla\upsilon
+\ldots
\nonumber\\
&&{}
+\frac{4i}{R}E^{\underline a}\,\upsilon\gamma_\star \Gamma^{b'}\Gamma_{\underline a}E\,\nabla\upsilon\Gamma_{b'}\Gamma_{11}E
-\frac{4i}{R}E^{\underline a}\,\upsilon\gamma_\star \Gamma^{b'}\Gamma_{\underline a}\Gamma_{11}E\,\nabla\upsilon\Gamma_{b'}E
\nonumber\\
&&{}
-\frac{2i}{R}E^{\underline a}\,E\Gamma_{11}E\,\upsilon\gamma_\star \Gamma_{\underline a}\nabla\upsilon
-\frac{2i}{R}E^{\underline a}\,E\gamma_7E\,\upsilon\gamma_\star \Gamma_{\underline a}\gamma_5\nabla\upsilon
\nonumber\\
&=&
\frac{4i}{R}E^{\underline a}\,\upsilon\gamma_\star \Gamma^{b'}\Gamma_{\underline a}E\,\nabla\upsilon\Gamma_{b'}\Gamma_{11}E
-\frac{4i}{R}E^{\underline a}\,\upsilon\gamma_\star \Gamma^{b'}\Gamma_{\underline a}\Gamma_{11}E\,\nabla\upsilon\Gamma_{b'}E
\nonumber\\
&&{}
-\frac{2i}{R}E^{\underline a}\,E\Gamma_{11}E\,\upsilon\gamma_\star \Gamma_{\underline a}\nabla\upsilon
-\frac{2i}{R}E^{\underline a}\,E\gamma_7E\,\upsilon\gamma_\star \Gamma_{\underline a}\gamma_5\nabla\upsilon\,,
\end{eqnarray}
where the first ellipsis denote the 5 terms which together with the previous term cancel due to the Fierz identity (\ref{eq:Fierz1}) and similarly for the second ellipsis. This leaves us with the following terms in $dH$
\begin{eqnarray}
&&{}
-\frac{2i}{R}E^B\,E\Gamma^AE\,\nabla\upsilon\Gamma_A\Gamma_{11}\Gamma_B\gamma_\star \upsilon
+\ldots
-\frac{i}{R}E^{b'}\,E\Gamma^AE\,\nabla\upsilon\Gamma_A\Gamma_{11}\Gamma_{b'}\gamma_\star \upsilon
+\ldots
\nonumber\\
&&{}
-\frac{i}{R}E^{b'}\,E\Gamma^AE\,\nabla\upsilon\gamma_\star \Gamma_{b'}\Gamma_A\Gamma_{11}\upsilon
+\ldots
+\frac{2i}{R}E^{b'}\,E\gamma_\star \Gamma_{b'}\Gamma^AE\,\nabla\upsilon\Gamma_A\Gamma_{11}\upsilon
+\ldots
\nonumber\\
&&{}
+\frac{4i}{R}E^{b'}\,E\gamma_\star \Gamma_{a'b'}\upsilon\,\nabla\upsilon\Gamma^{a'}\Gamma_{11}E
+\frac{4i}{R}E^{b'}\,E\gamma_\star \Gamma_{a'b'}\Gamma_{11}\upsilon\,\nabla\upsilon\Gamma^{a'}E
\nonumber\\
&&{}
+\frac{2i}{R}E^{a'}\,E\Gamma^{b\hat c}E\,\upsilon\Gamma_{a'b\hat c}\gamma_\star \Gamma_{11}\nabla\upsilon
-\frac{2i}{R}E^{b'}\,E\gamma_\star \Gamma_{11}E\,\nabla\upsilon\Gamma_{b'}\upsilon
-\frac{2i}{R}E^{b'}\,E\gamma E\,\nabla\upsilon\Gamma_{b'}\gamma_5\upsilon
\nonumber\\
&&{}
+\frac{2i}{R}E^{b'}\,E\Gamma_{11}E\,\upsilon\gamma_\star \Gamma_{b'}\nabla\upsilon
+\frac{2i}{R}E^{b'}\,E\gamma_7E\,\upsilon\gamma_\star \Gamma_{b'}\gamma_5\nabla\upsilon\,.
\end{eqnarray}
Again the ellipsis denotes terms which cancel together with the previous term due to (\ref{eq:Fierz1}). Using the relation
\begin{eqnarray}
\lefteqn{\frac{2i}{R}E^{a'}\,E\Gamma^{b\hat c}E\,\upsilon\Gamma_{a'b\hat c}\gamma_\star \Gamma_{11}\nabla\upsilon}
\nonumber\\
&=&
-\frac{i}{2R}E^{b'}\,E\gamma_\star \Gamma^{\underline{ca}}\Gamma_{11}E\,\nabla\upsilon\Gamma_{\underline{ac}b'}\upsilon
-\frac{i}{2R}E^{b'}\,E\gamma_\star \gamma_5\Gamma^{\underline{ca}}\Gamma_{11}E\,\nabla\upsilon\Gamma_{\underline{ac}b'}\gamma_5\upsilon
\nonumber\\
&=&
-\frac{i}{2R}E^{b'}\,E\gamma_\star \Gamma^{\underline c}\Gamma^A\Gamma_{11}E\,\nabla\upsilon\Gamma_A\Gamma_{\underline cb'}\upsilon
+\ldots
-\frac{i}{2R}E^{b'}\,E\gamma_\star \gamma_5\Gamma^{\underline c}\Gamma^A\Gamma_{11}E\,\nabla\upsilon\Gamma_A\Gamma_{\underline cb'}\gamma_5\upsilon
+\ldots
\nonumber\\
&&{}
-\frac{4i}{R}E^{b'}\,E\gamma_\star \Gamma_{a'b'}\upsilon\,\nabla\upsilon\Gamma^{a'}\Gamma_{11}E
-\frac{4i}{R}E^{b'}\,E\gamma_\star \Gamma_{a'b'}\Gamma_{11}\upsilon\,\nabla\upsilon\Gamma^{a'}E
\nonumber\\
&&{}
+\frac{2i}{R}E^{b'}\,E\gamma_7E\,\nabla\upsilon\Gamma_{b'}\gamma_\star \gamma_5\upsilon
+\frac{2i}{R}E^{b'}\,E\Gamma_{11}E\,\nabla\upsilon\Gamma_{b'}\gamma_\star \upsilon
\nonumber\\
&&{}
+\frac{2i}{R}E^{b'}\,E\gamma E\,\nabla\upsilon\Gamma_{b'}\gamma_5\upsilon
+\frac{2i}{R}E^{b'}\,E\gamma_\star \Gamma_{11}E\,\nabla\upsilon\Gamma_{b'}\upsilon
\nonumber\\
&=&
-\frac{4i}{R}E^{b'}\,E\gamma_\star \Gamma_{a'b'}\upsilon\,\nabla\upsilon\Gamma^{a'}\Gamma_{11}E
-\frac{4i}{R}E^{b'}\,E\gamma_\star \Gamma_{a'b'}\Gamma_{11}\upsilon\,\nabla\upsilon\Gamma^{a'}E
\nonumber\\
&&{}
+\frac{2i}{R}E^{b'}\,E\gamma_7E\,\nabla\upsilon\Gamma_{b'}\gamma_\star \gamma_5\upsilon
+\frac{2i}{R}E^{b'}\,E\Gamma_{11}E\,\nabla\upsilon\Gamma_{b'}\gamma_\star \upsilon
\nonumber\\
&&{}
+\frac{2i}{R}E^{b'}\,E\gamma E\,\nabla\upsilon\Gamma_{b'}\gamma_5\upsilon
+\frac{2i}{R}E^{b'}\,E\gamma_\star \Gamma_{11}E\,\nabla\upsilon\Gamma_{b'}\upsilon\,,
\end{eqnarray}
we see that also the last remaining seven terms in $dH$ cancel. This completes the proof that the NS--NS three-form we have constructed for $AdS_2\times S^2\times T^6$ is indeed closed.

\if{}
\bibliography{strings}

\begin{thebibliography}{10}

\bibitem{Sorokin:2010wn}
D.~Sorokin and L.~Wulff, ``{Evidence for the classical integrability of the
  complete AdS(4) x CP(3) superstring},''
  \href{http://dx.doi.org/10.1007/JHEP11(2010)143}{{\em JHEP} {\bfseries 11}
  (2010) 143},
\href{http://arxiv.org/abs/1009.3498}{{\ttfamily arXiv:1009.3498 [hep-th]}}.

\bibitem{Sorokin:2011rr}
D.~Sorokin, A.~Tseytlin, L.~Wulff, and K.~Zarembo, ``{Superstrings in
  $AdS(2)\times S(2) \times T(6)$},''
  \href{http://dx.doi.org/10.1088/1751-8113/44/27/275401}{{\em J.Phys.A}
  {\bfseries A44} (2011) 275401},
\href{http://arxiv.org/abs/1104.1793}{{\ttfamily arXiv:1104.1793 [hep-th]}}.

\bibitem{Bena:2003wd}
I.~Bena, J.~Polchinski, and R.~Roiban, ``{Hidden symmetries of the $AdS_5
  \times S^5$ superstring},''
  \href{http://dx.doi.org/10.1103/PhysRevD.69.046002}{{\em Phys. Rev.}
  {\bfseries D69} (2004) 046002},
\href{http://arxiv.org/abs/hep-th/0305116}{{\ttfamily arXiv:hep-th/0305116}}.

\bibitem{Adam:2007ws}
I.~Adam, A.~Dekel, L.~Mazzucato, and Y.~Oz, ``{Integrability of Type II
  Superstrings on Ramond-Ramond Backgrounds in Various Dimensions},''
  \href{http://dx.doi.org/10.1088/1126-6708/2007/06/085}{{\em JHEP} {\bfseries
  0706} (2007) 085}, \href{http://arxiv.org/abs/hep-th/0702083}{{\ttfamily
  arXiv:hep-th/0702083 [HEP-TH]}}.

\bibitem{Arutyunov:2008if}
G.~Arutyunov and S.~Frolov, ``{Superstrings on $AdS_4 \times CP^3$ as a Coset
  Sigma-model},'' \href{http://dx.doi.org/10.1088/1126-6708/2008/09/129}{{\em
  JHEP} {\bfseries 09} (2008) 129},
\href{http://arxiv.org/abs/0806.4940}{{\ttfamily arXiv:0806.4940 [hep-th]}}.

\bibitem{Stefanski:2008ik}
B.~Stefanski~Jr., ``{Green-Schwarz action for Type IIA strings on $AdS_4\times
  CP^3$},'' \href{http://dx.doi.org/10.1016/j.nuclphysb.2008.09.015}{{\em Nucl.
  Phys.} {\bfseries B808} (2009) 80--87},
\href{http://arxiv.org/abs/0806.4948}{{\ttfamily arXiv:0806.4948 [hep-th]}}.

\bibitem{Gomis:2008jt}
J.~Gomis, D.~Sorokin, and L.~Wulff, ``{The complete AdS(4) x CP(3) superspace
  for the type IIA superstring and D-branes},''
  \href{http://dx.doi.org/10.1088/1126-6708/2009/03/015}{{\em JHEP} {\bfseries
  03} (2009) 015},
\href{http://arxiv.org/abs/0811.1566}{{\ttfamily arXiv:0811.1566 [hep-th]}}.

\bibitem{Babichenko:2009dk}
A.~Babichenko, B.~Stefanski, Jr., and K.~Zarembo, ``{Integrability and the
  AdS(3)/CFT(2) correspondence},''
  \href{http://dx.doi.org/10.1007/JHEP03(2010)058}{{\em JHEP} {\bfseries 03}
  (2010) 058},
\href{http://arxiv.org/abs/0912.1723}{{\ttfamily arXiv:0912.1723 [hep-th]}}.

\bibitem{Cagnazzo:2009zh}
A.~Cagnazzo, D.~Sorokin, and L.~Wulff, ``{String instanton in AdS(4)xCP(3)},''
  \href{http://dx.doi.org/10.1007/JHEP05(2010)009}{{\em JHEP} {\bfseries 05}
  (2010) 009},
\href{http://arxiv.org/abs/0911.5228}{{\ttfamily arXiv:0911.5228 [hep-th]}}.

\bibitem{Astolfi:2009qh}
D.~Astolfi, V.~G.~M. Puletti, G.~Grignani, T.~Harmark, and M.~Orselli, ``{Full
  Lagrangian and Hamiltonian for quantum strings on $AdS_4 \times CP^3$ in a
  near plane wave limit},''
  \href{http://dx.doi.org/10.1007/JHEP04(2010)079}{{\em JHEP} {\bfseries 1004}
  (2010) 079}, \href{http://arxiv.org/abs/0912.2257}{{\ttfamily arXiv:0912.2257
  [hep-th]}}.

\bibitem{Grassi:2009yj}
P.~A. Grassi, D.~Sorokin, and L.~Wulff, ``{Simplifying superstring and D-brane
  actions in AdS(4) x CP(3) superbackground},''
  \href{http://dx.doi.org/10.1088/1126-6708/2009/08/060}{{\em JHEP} {\bfseries
  08} (2009) 060},
\href{http://arxiv.org/abs/0903.5407}{{\ttfamily arXiv:0903.5407 [hep-th]}}.

\bibitem{Beisert:2005bm}
N.~Beisert, V.~Kazakov, K.~Sakai, and K.~Zarembo, ``{The Algebraic curve of
  classical superstrings on $AdS_5 \times S^5$},''
  \href{http://dx.doi.org/10.1007/s00220-006-1529-4}{{\em Commun.Math.Phys.}
  {\bfseries 263} (2006) 659--710},
\href{http://arxiv.org/abs/hep-th/0502226}{{\ttfamily arXiv:hep-th/0502226
  [hep-th]}}.

\bibitem{Gromov:2008bz}
N.~Gromov and P.~Vieira, ``{The AdS4/CFT3 algebraic curve},''
  \href{http://dx.doi.org/10.1088/1126-6708/2009/02/040}{{\em JHEP} {\bfseries
  02} (2009) 040},
\href{http://arxiv.org/abs/0807.0437}{{\ttfamily arXiv:0807.0437 [hep-th]}}.

\bibitem{SchaferNameki:2010jy}
S.~Schafer-Nameki, ``{Review of AdS/CFT Integrability, Chapter II.4: The
  Spectral Curve},''
\href{http://arxiv.org/abs/1012.3989}{{\ttfamily arXiv:1012.3989 [hep-th]}}.

\bibitem{Howe:2004ib}
P.~S. Howe and E.~Sezgin, ``{The supermembrane revisited},''
  \href{http://dx.doi.org/10.1088/0264-9381/22/11/017}{{\em Class. Quant.
  Grav.} {\bfseries 22} (2005) 2167--2200},
\href{http://arxiv.org/abs/hep-th/0412245}{{\ttfamily arXiv:hep-th/0412245}}.

\bibitem{Grisaru:1985fv}
M.~T. Grisaru, P.~S. Howe, L.~Mezincescu, B.~Nilsson, and P.~Townsend, ``{N=2
  Superstrings in a Supergravity Background},''
  \href{http://dx.doi.org/10.1016/0370-2693(85)91071-8}{{\em Phys.Lett.}
  {\bfseries B162} (1985) 116}.

\bibitem{Berkovits:2001ue}
N.~Berkovits and P.~S. Howe, ``{Ten-dimensional supergravity constraints from
  the pure spinor formalism for the superstring},''
  \href{http://dx.doi.org/10.1016/S0550-3213(02)00352-8}{{\em Nucl.Phys.}
  {\bfseries B635} (2002) 75--105},
  \href{http://arxiv.org/abs/hep-th/0112160}{{\ttfamily arXiv:hep-th/0112160
  [hep-th]}}.

\bibitem{Fre:2008qc}
P.~Fr\'e and P.~A. Grassi, ``{Pure Spinor Formalism for ${Osp}(N|4)$
  backgrounds},''
\href{http://arxiv.org/abs/0807.0044}{{\ttfamily arXiv:0807.0044 [hep-th]}}.

\bibitem{Bonelli:2008us}
G.~Bonelli, P.~A. Grassi, and H.~Safaai, ``{Exploring Pure Spinor String Theory
  on $AdS_4\times \mathbf{CP}^3$},''
  \href{http://dx.doi.org/10.1088/1126-6708/2008/10/085}{{\em JHEP} {\bfseries
  10} (2008) 085},
\href{http://arxiv.org/abs/0808.1051}{{\ttfamily arXiv:0808.1051 [hep-th]}}.

\bibitem{D'Auria:2008cw}
R.~D'Auria, P.~Fre, P.~A. Grassi, and M.~Trigiante, ``{Superstrings on
  $AdS_4\times CP^3$ from Supergravity},'' {\em Phys. Rev.} {\bfseries D79}
  (2009) 086001,
\href{http://arxiv.org/abs/0808.1282}{{\ttfamily arXiv:0808.1282 [hep-th]}}.


\end{thebibliography}
\bibliographystyle{utphys}
\end{document}
\fi

\providecommand{\href}[2]{#2}\begingroup\raggedright\endgroup

\end{document}